\documentclass[twocolumn]{aastex631}

\usepackage{soul}
\usepackage{amsmath}
\usepackage{afterpage}
\usepackage{tabularx}
\usepackage{ltablex}
\usepackage{ulem} 
\usepackage{gensymb}
\usepackage{cleveref}
\usepackage{hyperref}

\newcommand{\simname}[1]{\texttt{#1}}
\newcommand{\orcid}[1]{\href{https://orcid.org/#1}{\includegraphics[width=10pt]{orcid_id.png}}}

\let\oldhat\hat
\renewcommand{\vec}[1]{\boldsymbol{#1}} 
\renewcommand{\hat}[1]{\oldhat{\boldsymbol{#1}}}

\newcommand{\beq}{\begin{equation}}
\newcommand{\eeq}{\end{equation}}

\newcommand{\sigmadustg}{\Sigma_{\rm d,g}}

\newcommand{\Temp}{T_{\rm mp}}
\newcommand{\Msun}{M_{\odot}}

\newcommand{\upmu}{\rm \mu}

\newcommand{\mearth}{M_{\oplus}}

\received{October 8, 2024}
\accepted{February 19, 2025}
\published{XXX}

\shorttitle{Accretion bursts in young intermediate-mass stars (YIMSs)}
\shortauthors{Das et al. (2025)}

\begin{document}

\title{Accretion bursts in young intermediate-mass stars make planet formation challenging}

\correspondingauthor{Indrani Das, Eduard Vorobyov, Shantanu Basu}
\email{idas2@uwo.ca, eduard.vorobiev@univie.ac.at, basu@uwo.ca}

\author[0000-0002-7424-4193]{Indrani Das}
\affiliation{Institute of Astronomy and Astrophysics, Academia Sinica, No. 1, Sec. 4, Roosevelt Road, Taipei 10617, Taiwan}

\author[0000-0002-6045-0359]{Eduard Vorobyov}
\affiliation{
     Ural Federal University, 19 Mira Str., 620002 Ekaterinburg, Russia}
\affiliation{Research Institute of Physics, Southern Federal University, Rostov-on-Don 344090, Russia }

\author[0000-0003-0855-350X]{Shantanu Basu}
\affiliation{University of Western Ontario, Department of Physics and Astronomy, London, ON N6A 3K7, Canada}
\affiliation{Canadian Institute for Theoretical Astrophysics, University of Toronto, 60 St. George, St., Toronto, ON M5S 3H8, Canada}

\begin{abstract}

We investigate the occurrence of accretion bursts, dust accumulation, and the prospects for planetesimal formation in a gravitationally unstable magnetized protoplanetary disk with globally suppressed but episodically triggered magnetorotational instability (MRI), particularly in young intermediate-mass stars (YIMSs) but with a brief comparison to low-mass counterparts.
We use numerical magnetohydrodynamics simulations in the thin-disk limit (FEOSAD code) to model the formation and long-term evolution of a gravitationally unstable magnetized protoplanetary disk, including dust dynamics and growth, since the collapse of a massive slowly-rotating prestellar cloud core.
Massive gas concentrations and dust rings  form within the inner disk region owing to the radially varying efficiency of mass transport by gravitational instability (GI). These rings are initially susceptible to  streaming instability (SI). However, gradual warming of the dust rings thanks to high opacity and GI-induced influx of matter increases the gas temperature above a threshold for the MRI to develop via thermal ionization of alkaline metals. The ensuing MRI bursts destroy the dust rings, making planetesimal formation via SI problematic. In the later evolution phase, when the burst activity starts to diminish, SI becomes inefficient because of growing dust drift velocity and more extended inner dead zone, both acting to reduce the dust concentration below the threshold for SI to develop. Low-mass objects appear to be less affected by these adverse effects.
Our results suggest that disks around young intermediate-mass stars may be challenging environments for planetesimal formation via SI. This may explain the dearth of planets around stars with $M_\ast > 3.0$~$M_\odot$.
\end{abstract}


\section{Introduction} 
The young intermediate-mass stars (YIMSs), spanning a mass range of about 2 to 10 $\Msun$, bridge the gap between low-mass stars and high-mass stars. 
YIMSs have some similar properties as a classical T Tauri stars, for instance having emission lines, UV-excess, a lower surface gravity than main-sequence stars \citep{HamannPersson1992,Vink+2005} and they are usually identified by an infrared excess 
from circumstellar disks \citep{vandenAncker+2000,Meeus+2001}. 
With Gaia-updated
distances, using the ALMA archival data for a Herbig disk population study, \cite{Stapper+2022} find that Herbig disks are on average 3-7 times more massive than the typical disk around a T Tauri star, and also likely larger.

The mass accretion rate is a key
physical quantity in star formation to 
determine the process of mass assembly and the associated timescale that results in the formation of a protostar, and can be used to infer the total stellar luminosity across
a wide mass range of various stellar objects: from classical T-Tauri stars to YIMSs \citep{Fairlamb+2015,Manara+2016}. 
Understanding the accretion signatures enhances our knowledge of YIMS formation and comparison to their low-mass counterparts. 
In a spectroscopic study of 163 Herbig Ae/Be stars complemented with the X-Shooter sample and updated with Gaia DR2 parallaxes, \cite{Wichittanakom+2020} determine mass accretion rates ranging from about $10^{-7.79}$ to $10^{-3.61}~M_\odot$~yr$^{-1}$ as derived from the ${\rm H \alpha}$ line emission for the accretion luminosity going from about $10^{-0.35}$ to $10^{4.08}~L_\odot$. 
While for T Tauri stars the  observed mass accretion rate ranges from roughly about $10^{-8}$ to $10^{-4}~M_\odot$~yr$^{-1}$ corresponding to the accretion luminosity going from about $1$ to $10^{3}~L_\odot$ \citep[e.g.,][and references therein]{WhiteBasri2003, Calvet+2004,Natta+2006,Cieza+2018,
Perez+2020,Fischer+2023,Senra+2024}.

The dependence of planetary mass on stellar mass as well as robust statistical evidence of disk evolution shows an
apparent paucity of planets around intermediate-mass stars \citep{Ribas+2015,Janson+2021,DelgadoMena+2023,Ronco+2024}. 
On the other hand, the high occurrence of exoplanets discovered around a host star with masses $< 2.5$~$\Msun$ indicates greater feasibility of planetary formation around low-mass stars. 
Therefore, YIMSs are considered to be prime laboratories for studying the physical processes that may inhibit or alter planet formation, potentially explaining the dearth of planets around host stars with masses more massive than $\sim2.5-3.0$~$M_{\odot}$  
as found from the radial velocities and direct imaging observations \citep[][see further in \href{https://exoplanets.eu/}{exoplanets.eu}]{Johnson+2010PASP,Reffert+2015}. 
The theoretical studies by \cite{Ida+2008,Kennedykenyon2008ApJ} show that planetary formation may also decrease around stars with masses greater than $2-3$~$\Msun$. 
This occurs because the water snowline -- a major site of planetesimal formation -- is pushed with the increasing stellar luminosity further out to the disk regions where the  timescale of planetesimal formation is longer than the timescale of protoplanetary core migration. 
A recent study by \cite{Pinilla+2022} also shows the significance of the radial dust drift velocity becoming a potential barrier for planetesimal formation around a host star with masses $M_* > 1.5$~$\Msun$. 
Therefore, it is crucial to study the long-term evolution of the protoplanetary disk around such a YIMS that may help explain why planets around YIMSs may be rare.

The main aim of this paper is to determine the episodic accretion history, dust accumulation,
and prospects for planetesimal formation in the inner regions of a 
protoplanetary disk (PPD) around a YIMS since the embedded phase. 
We investigate the formation and evolution of the PPD  
with long-term global magnetohydrodynamic simulations that have dynamically coupled dust and gas and also account for self-gravity and the triggering of MRI in an inner region of the PPD.
In the early embedded stages of disk evolution, the accumulation of gas and dust in the dead zones (inner disk regions characterized by a reduced rate of mass transport and low ionization) leads to disk instabilities and GI-driven MRI outbursts.  
We update our understanding of the accretion outbursts of YIMSs by comparative analysis of their characteristics with respect to their low-mass cousins. 
We also investigate the prospects of first-generation planetesimal formation within the massive dusty rings forming within the innermost few au of the PPD via streaming instability (SI), and the consequences of episodic accretion that can explain the paucity of planets around stars with masses $M_\ast > 3.0$~$M_\odot$.

The structure of this paper is as follows. 
Section \ref{sec:methodpaper} outlines our numerical setup. 
Section \ref{sec:results} presents an analysis of our findings, with Section \ref{sec:results1} covering the episodic accretion history of a young intermediate-mass star and its PPD evolution, and Section \ref{sec:SI} discussing the prospects of streaming instability within the PPD. 
Section \ref{sec:discussion} discusses our results and their significance in the context of contemporary observations. 
Finally, Section \ref{sec:conclusions} summarizes the main conclusions of our study.

\section{Numerical Methods} \label{sec:methodpaper}

We perform our numerical global magnetohydrodynamic simulations starting from the gravitational collapse of a prestellar magnetized cloud core using the code Formation and Evolution Of a Star And its circumstellar Disk (FEOSAD) \citep{Vorobyov+2020} to study the formation and long-term evolution (up to 1 Myr) of PPDs in intermediate-mass stars. 
Such long integration times are made possible by the use of the thin-disk approximation, the justification of which is
provided in \cite{VorobyovBasu2010}. 
In our model, the prestellar core has the form of a flattened pseudo-disk, a spatial configuration
that can be expected in the presence of rotation and large-scale magnetic fields \citep{Basu1997}. As the collapse proceeds, the inner regions of the core spin up and a centrifugally supported disk forms when the inner infalling layers of the core hit the centrifugal barrier near the central sink cell at ${r_{\rm sc}} =0.52 \, {\rm au}$. The evolution gradually continues into the embedded phase of star formation where the nascent central star is surrounded by a rotationally supported accretion disk along with remaining infalling envelope, followed by the Herbig stage wherein the envelope diminishes and the central star attains its terminal mass of about 4.5~$M_\odot$.
In our numerical study, such a choice for the final mass of the host star helps in assessing the significant deviations in the PPD evolution and dust dynamics from the low-mass counterparts, 
which can otherwise be less pronounced at about 3~$\Msun$, given that this boundary may be biased due to limitations in observational constraints.

FEOSAD takes into account the co-evolution of the gas and dust components, including dust growth and back-reaction of dust on gas. Key thermal processes, such as stellar irradiation, viscous heating, dust thermal cooling, and $PdV$ work, are considered to provide realistic temperature structure of the disk. Disk self-gravity is computed using the integral form of the gravitational potential \citep{Vorobyov+2024}.  
The effects of magnetic fields are accounted for approximately, assuming the flux-freezing approximation and a simplified magnetic field geometry appropriate for the thin-disk approximation \citep{Basu1997}. Considering magnetic fields allows us to construct a more sophisticated model of the MRI bursts than the usual layered-disk model approximation \citep{Gammie1996}. The PPD undergoes MRI outbursts in the inner several astronomical units when forming and evolving, while disk dynamics in the intermediate and outer regions is being played out by gravitational instability. 
The initial background turbulent viscosity across the disk extent is set equal an equivalent of $\alpha_{\rm visc}=10^{-4}$, in accordance with recent observations of efficient dust settling in protoplanetary disks \citep{Rosotti2023}. 
In the MRI-dead layer, a viscosity parameter of $\alpha_{\rm dz}=10^{-5}$ is set due to nonzero residual transport, which may be caused by low-level turbulence propagating from the upper MRI-active disk layers down the disk midplane.
Heating of the dead zone owing to residual turbulence and PdV work can raise the gas temperature above 1000~K. 
The thermal ionization of the alkaline metals enables the vigorous MRI growth across the dead zone, followed by a rapid transport of the inner disk material (which is associated with a turbulent viscosity of $\alpha_{\rm visc}=0.1$ in our numerical model, see further in Appendix \ref{sec:viscosity_ion}) that gives rise to the MRI-triggered accretion
bursts \citep[][]{Armitage+2001,Zhu+2009,Vorobyov+2020,
Kadam+2022}.

We refer to Section \ref{sec:appmethods} for the detailed information on the gas and dust physics (Sec \ref{sec:gas} and Sec \ref{sec:dust}), numerical treatment of the adaptive viscosity and ionization fraction during the MRI bursts (Sec \ref{sec:viscosity_ion}), initial and boundary conditions (Sec \ref{sec:ICs}). The details of the model parameters of our conducted numerical simulations are listed in Table \ref{tab:table}.

\begin{table*}[!ht]
\centering
\begin{tabular}{ccccccccccccc}
    \hline \\
    Model & $M_{\rm core}$ & $r_{\rm out}$ & $r_{\rm in}$ & $r_0$ & $\Sigma_{g,0}$ & $\Omega_0$  &  $\beta$ [\%] & $\lambda$ & 
   $\alpha_{\rm MRI}^{\rm}$ & $\alpha_{\rm MRI}^{\rm Burst}$ 
    & $\alpha_{\rm dz}^{\rm}$ 
    & $M_{*,{\rm f}}$ \\ 
    & [$M_{\odot}$] &  [au] &  [au] &  [au] & [g.${\rm cm}^{-2}$]  & $[{\rm km} \ {\rm /s /pc}]$ & & & & & &[$M_{\odot}$] \\ 
    \hline \\
     \simname{model-YIMS} & 5.99 & $3.61\times10^{4}$ & 0.52 & $3.73\times10^{2}$ & $8.58\times10^{-2}$ & 0.80  & $2.14\times10^{-4}$ & 10.0 & $10^{-4}$ & 0.10 & $10^{-5}$ & 4.50\\
     \simname{model-LMS} & 1.01 & $5.05\times10^{3}$ & 0.52 & $6.30\times10^{2}$ & $5.08\times10^{-1}$ & 2.00  & $1.33\times10^{-3}$ & 10.0 & $10^{-4}$ & 0.10 & $10^{-5}$ & 0.71\\
     \hline \\
\end{tabular}
\caption{Model parameters (from left to right): $M_{\rm core}$ is the initial core mass, $r_{\rm out}$ is the initial outer radius of the prestellar core, ${\rm r_{in}}$ is inner boundary of the computational domain, $r_0$ is the radius of the central plateau in the initial core, $\Sigma_{\rm g,0}$ is the initial gas surface density at the center of the core , 
$\Omega_0$ is initial angular speed of the prestellar rotating core, $\beta$ is initial ratio of the rotational to gravitational energy, $\lambda$ is the initial mass to magnetic-flux ratio normalized to a factor of $(2\pi\sqrt{G})^{-1}$,   
$\alpha_{\rm MRI}$ is the constant background turbulent viscosity across the disk extent, $\alpha_{\rm MRI}^{\rm Burst}$ is the maximum turbulent viscosity at the occurrence of an MRI outburst only, $\alpha_{\rm dz}$ is the nonzero residual viscosity parameter in the MRI-dead layers in the absence of the burst,   
$M_{*,{\rm f}}$ is the final stellar mass at a desired timeline of $\sim\,1.0 \ {\rm Myr}$.} 
\label{tab:table}
\end{table*}

\section{Results} \label{sec:results}

In this section, we delineate our comprehensive findings on the general characteristics of GI-controlled disks around YIMSs, with a focus on how these characteristics evolve with the onset and decay of an accretion burst. 
Thereafter, we delve into the evolution and growth of dust, exploring the development of SI and its implications for planetesimal formation in PPDs around YIMSs but with a brief comparison to low-mass counterparts.

\subsection{Accretion History during Disk Evolution} \label{sec:results1} 
\begin{figure*}[htb!]
\hspace*{-1.5cm}
\includegraphics[width=22cm, height=7cm]{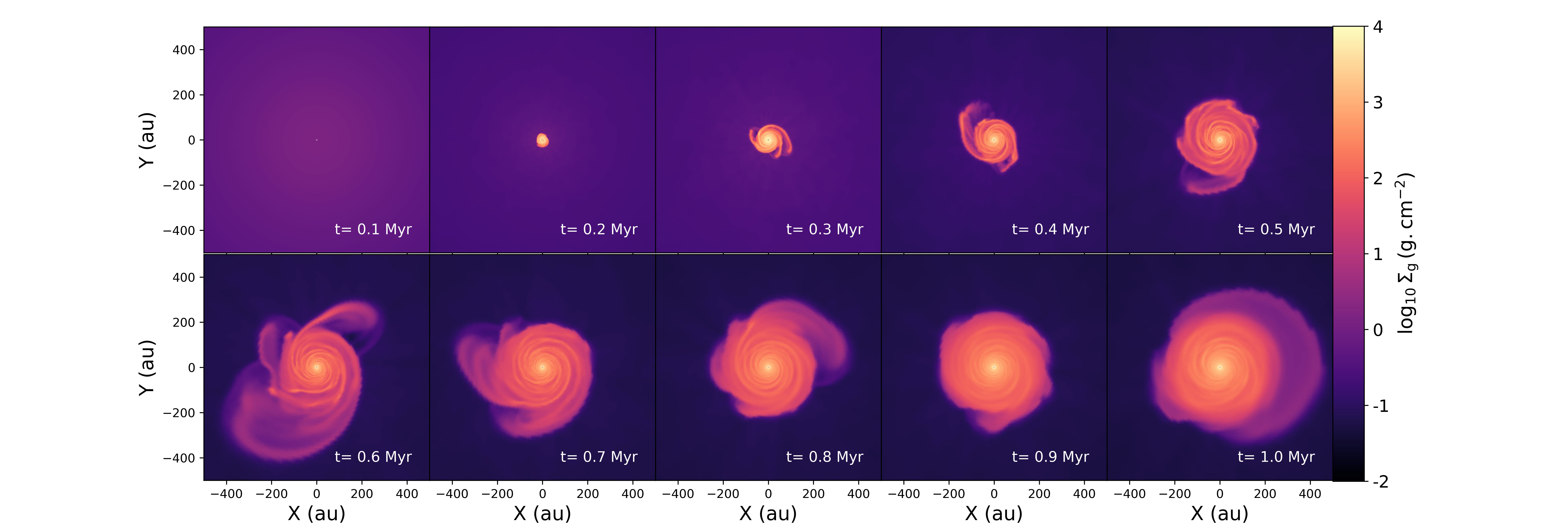}
\caption{Evolution of the gas surface density distribution over a region of $500 \times 500 \, {\rm au}^2$ in the midplane of the disk (in units of $\log_{10} \, {\rm g \, {cm}^{-2}}$), showing the large-scale disk structure at distinct time instances after the onset of collapse as obtained from \simname{model-YIMS}}. 
\label{fig:2DEvolDisk}
\end{figure*}

\begin{figure*}[htb!]
    \begin{minipage}{\linewidth}
    \hspace*{-1.5cm}
      \includegraphics[width=22cm, height=7cm]{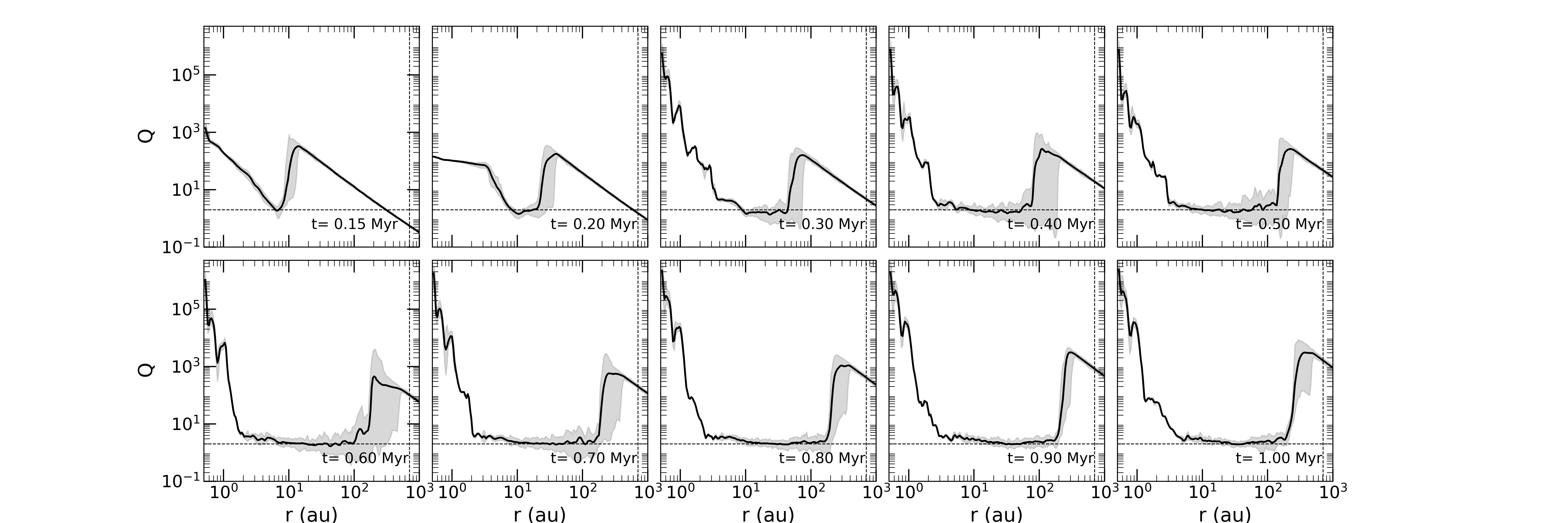}
    \end{minipage}
\caption{Radial distribution of the Toomre $Q$ parameter over a region of 1000~au, showing the azimuthally-averaged values (thick black lines) and the azimuthal scatter at a given radial distance (gray shaded area) as obtained from \simname{model-YIMS}. The horizontal dashed line corresponds to $Q=2$.} 
\label{fig:2DQparDisk}
\end{figure*}

\begin{figure}[htb!]
\epsscale{1}
\plottwo{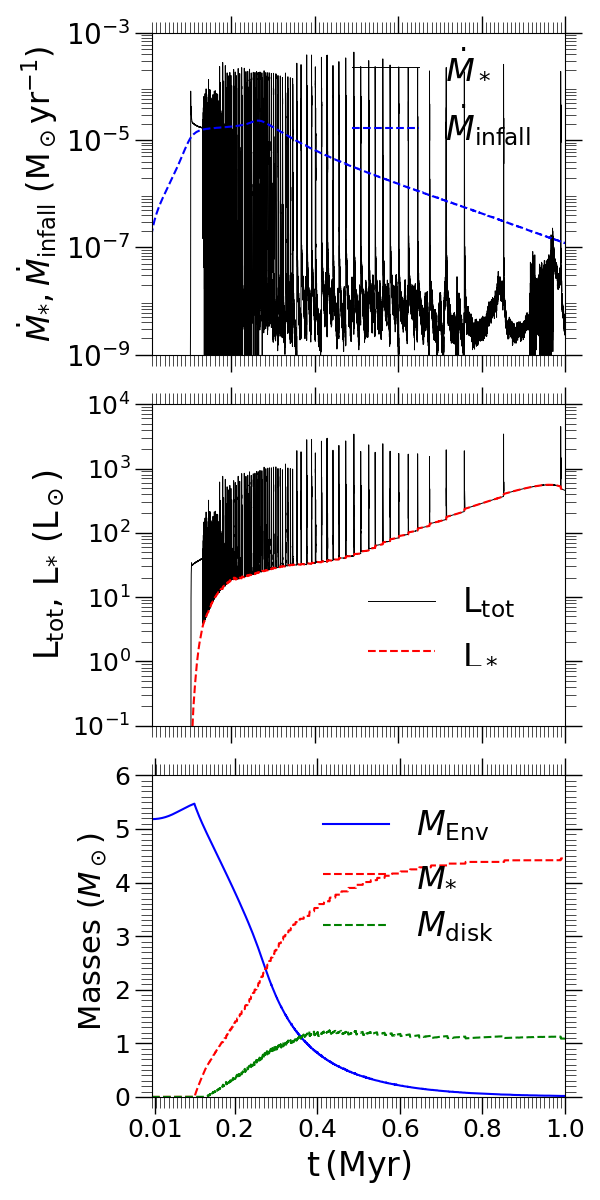}{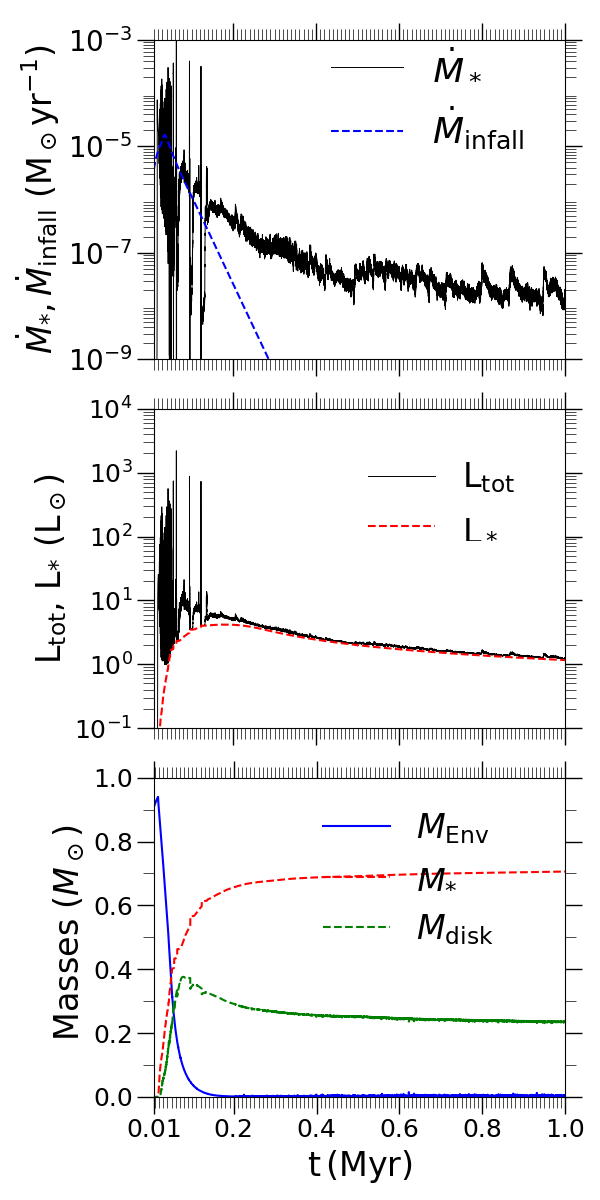}
\caption{Temporal evolution of disk accretion quantities at the sink-disk interface for \simname{model-YIMS} (left figure) and \simname{model-LMS} (right figure). 
Top panel: Mass accretion rate onto the star ($\dot{M}_{\ast}$) as shown by the black curve along with the mass infall rate from envelope ($\dot{M}_{\rm infall}$) as shown by the blue dashed curve. Middle panel: Total luminosity ($L_{\rm tot}$) as shown by the black curve combining the accretion luminosity ($L_{\rm acc}$) and the photospheric luminosity ($L_{\ast}$). Here, red dashed curve shows the photospheric luminosity.
Bottom panel: Time evolution of the envelope mass, star mass, and disk mass.} 
\label{fig:1DEvolDiskV2}
\end{figure}

\begin{figure*}[htb!]
    \centering
    \begin{minipage}{\linewidth}
        \centering
\includegraphics[width=\linewidth]{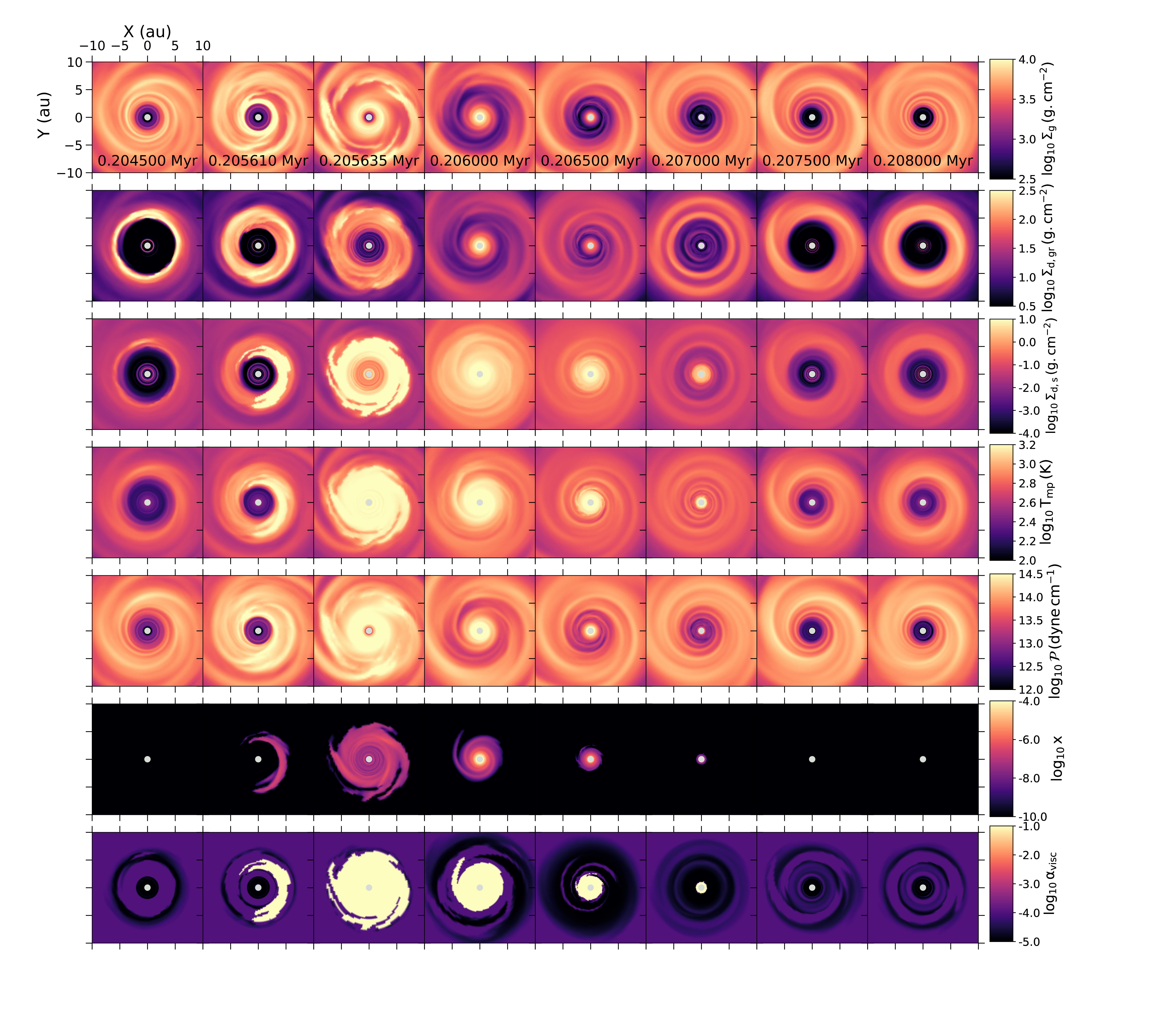}
    \end{minipage}
    \hfill
    \begin{minipage}{\linewidth}
    \vspace{-1cm}
         \hspace{-1cm}
        \centering        \includegraphics[width=0.8\linewidth, height=4cm]{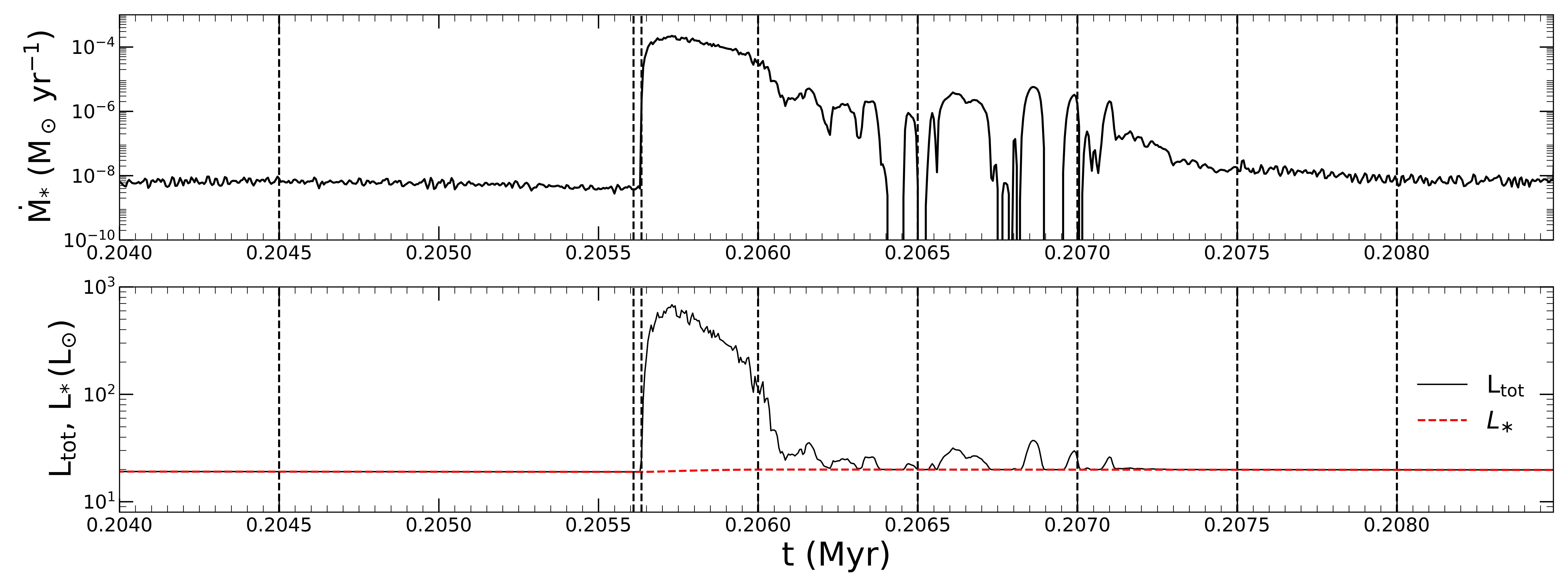}
    \end{minipage}
    \hfill
\caption{The evolution of an outside-in MRI outburst as it rises and decays back to the quiescent phase as presented in the intensity maps for \simname{model-YIMS}. Evolution of the gas surface density $\Sigma_{\rm g}$ (in units of $\log_{10} \, {\rm g \, {cm}^{-2}}$), grown dust surface density $\Sigma_{\rm d,gr}$ (in units of $\log_{10} \, {\rm g \, {cm}^{-2}}$), small dust surface density $\Sigma_{\rm d,s}$ (in units of $\log_{10} \, {\rm g \, {cm}^{-2}}$), midplane temperature $\Temp$ (in units of ${\rm K}$), 
vertically integrated midplane pressure $\mathcal{P}$ (in units of $\log_{10} \, {\rm dyne \, {cm}^{-1}}$), 
ionization-fraction ($x$), and turbulent viscosity ($\alpha_{\rm visc}$) distribution over a region of $20 \times 20 \, {\rm au}^2$ in the midplane of the disk. Last two panels present the temporal evolution of mass accretion rate onto the star (${\dot{M}}_\ast$) and total luminosity ($ L_{\rm tot}=L_{\rm acc}+L_{\ast}$) along with the contribution from the photospheric luminosity ($L_{\ast}$) at the sink-disk interface as an outburst develops and decays, obtained from \simname{model-YIMS}. The vertical dashed lines mark the time instances corresponding to those of the above 2D intensity maps of the disk characteristics.} 
\label{fig:gasdust2DEvol}
\end{figure*}

\begin{figure*}[!htb]
\vspace*{-0.2cm}
\gridline{\fig{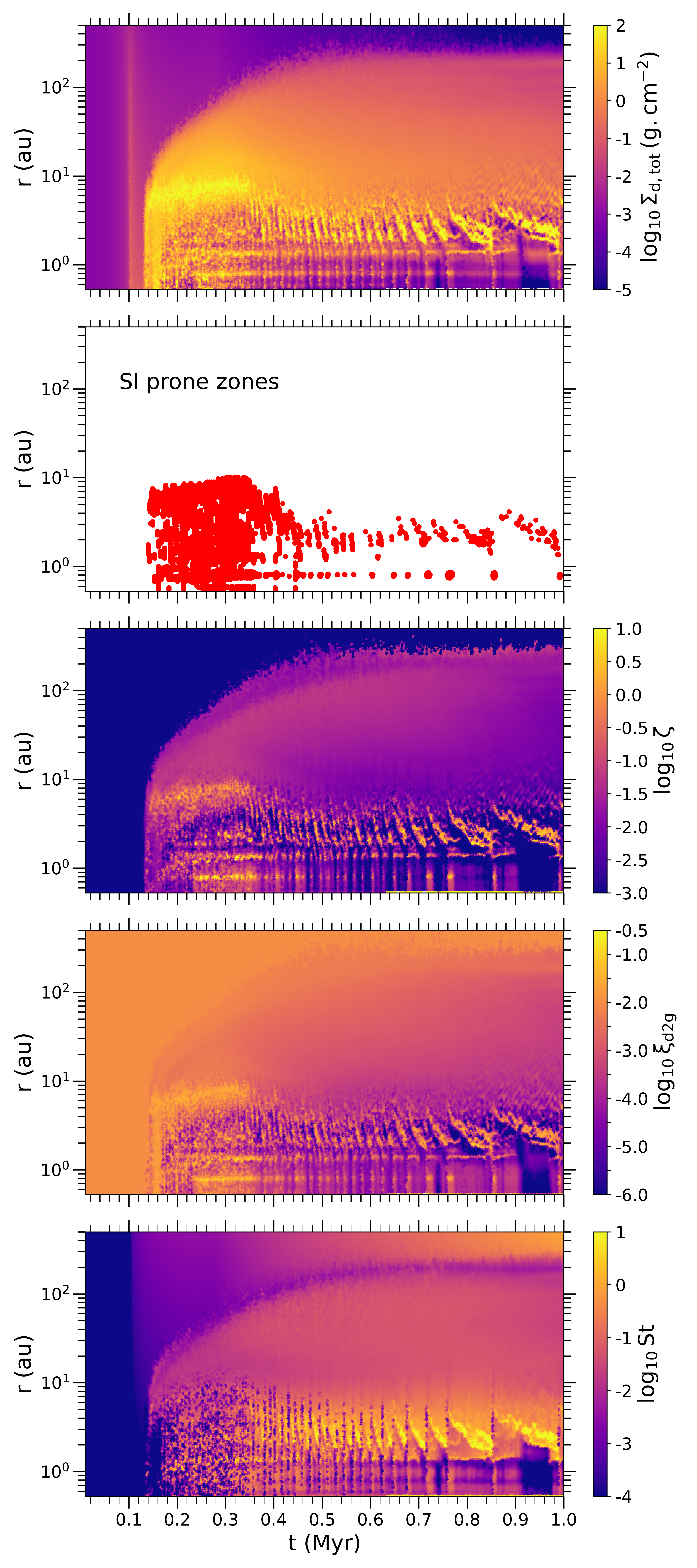}{0.54\textwidth}{(a)}
\hspace*{-0.4cm}
\fig{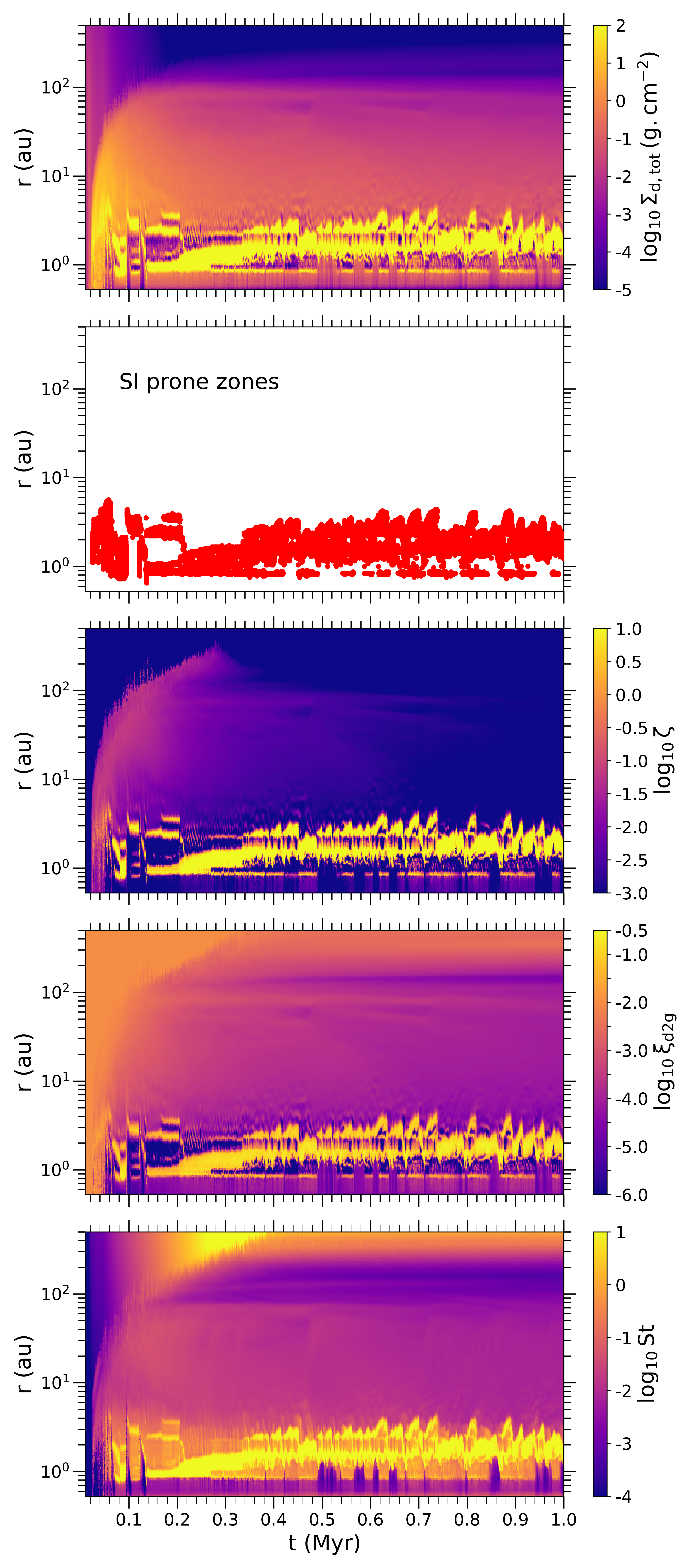}{0.54\textwidth}{(b)}}
\vspace*{-0.4cm}
\caption{Development of SI-prone zones within the PPD as seen from the spacetime profiles of total dust surface density $\Sigma_{\rm d,tot}$, the region in red presenting the streaming instability (SI) prone zones within the dust rings (where both the \cite{Yang+2017}-criterion as well as \cite{YoudinGoodman2005}-criterion are satisfied), ratio of volume densities of grown dust to gas $\zeta= \rho_{\rm d, gr}/\rho_{\rm g}$, dust-to-gas mass ratio ($\xi_{\rm d2g}= \Sigma_{\rm d,tot}/\Sigma_{\rm g}$), and Stokes number ${\rm St}$. 
Left (a) and right (b) figure show the cases for \simname{model-YIMS} and  \simname{model-LMS}, respectively.}
\label{fig:SISpacetime}
\end{figure*}

Figure \ref{fig:2DEvolDisk} shows the global view of the long-term evolution of a protoplanetary disk since the gravitational collapse of a massive prestellar core of $6.0~M_\odot$, where the gas surface density is depicted over a region of 500 au $\times$ 500 au. 
The disk forms at $t \approx 0.098 \,{\rm Myr}$ after the onset of collapse. 
In the first phase, by $t\lesssim 0.5\,{\rm Myr}$, the disk quickly grows up to a radius of $\approx 200 \, {\rm au}$. The disk growth is predominantly regulated by mass gain from the collapsing cloud and mass loss via protostellar accretion. For a globally low $\alpha_{\rm visc}=10^{-4}$, the inward mass transport across the disk is mostly provided by vigorous disk GI \citep{VorobyovBasu2009}, which is manifested by the appearance of irregular spiral structures. By $t \approx 0.5$~Myr, the mass of the star reaches $4.0~M_\odot$ and the disk mass is about 1.2~$M_\odot$, whereas less than $0.5~M_\odot$ is left in the infalling envelope. 
Later on, when the mass infall diminishes, the disk radius slowly increases with time and infrequently contains grand spiral arms. In this later phase, the disk mass slowly decreases and the lack of compensating infall leads to weakening of GI. 
The disk slowly spreads out as the gravitational torques in the outer disk tend to be positive \citep{VorobyovBasu2007}. The analysis of the Toomre $Q$-parameter confirms the susceptibility of the disk to GI and its gradual stabilization with time (see Fig.~\ref{fig:2DQparDisk}). The stellar mass by the end of simulations reaches $4.5~M_\odot$ and the envelope is completely depleted.

Figure~\ref{fig:2DQparDisk} presents the distribution of the Toomre $Q$-parameter in the disk during the entire evolution period considered. Several trends are to be emphasized in the context of our work. 
First, the gravitationally unstable region of the disk rapidly grows with time from a narrow annulus around 10-20 au at about 0.15~Myr to a wide region, ranging from several astronomical units to about 120~au by about 0.5~Myr. 
Afterwards, by 1 Myr, the GI unstable region in the outer disk further expands to about 200~au. 
Second, on both sides of this region $Q$ sharply increases. The increase beyond 100~au is related to the transition from the disk to rarefied envelope. The rise in the inner several astronomical units is caused by disk warming and high rates of shear. This latter feature is important and causes a bottleneck effect when material is transported by gravitational torques from outer to inner disk regions \citep[see][for details]{Vorobyov+2024}. The accumulation of gas and dust in this inner region form a dead zone, which is episodically activated by the development of the MRI owing to thermal ionization of alkaline metals (see Sect.~\ref{sec:results1}). 
The streaming instability can
be triggered within this dead zone (see Sect.~\ref{sec:SI}).

Figure \ref{fig:1DEvolDiskV2} presents the temporal evolution of the mass accretion rate along with total luminosity and the mass evolution of the envelope, disk and star for a YIMS and a low-mass star (hereafter LMS).
For YIMS, the mass accretion history is characterized by frequent accretion bursts with magnitude $\ge 10^{-4}~M_\odot$~yr$^{-1}$ superimposed on the quiescent rate of $\approx 10^{-8}~M_\odot$~yr$^{-1}$.
The accretion bursts are more frequent in the early stages of evolution and persist even at later times up to 0.85~Myr. 
Whereas, for the case of LMS, mass accretion history is characterized by a smaller number of accretion bursts with magnitude $\ge 5\times10^{-5}~M_\odot$~yr$^{-1}$ superimposed on the quiescent rate of $\approx 10^{-6}~M_\odot$~yr$^{-1}$ at about 0.1~Myr, gradually declining to $\approx 10^{-7}~M_\odot$~yr$^{-1}$ by 0.5~Myr, then saturating at $\approx 10^{-8}~M_\odot$~yr$^{-1}$ by 1.0~Myr. 
A correlation between the mass infall rate and the frequency of the bursts is also evident, since the mass infall is an important  part of the burst mechanism. In our model, the bursts are caused by the MRI development in the innermost disk regions, wherein slow disk heating causes thermal ionization of alkaline metals. Because each burst consumes about $0.04-0.1\, M_\odot$ of the disk mass for \simname{model-YIMS} and $0.004-0.05\, M_\odot$ of the disk mass for  \simname{model-LMS}, respectively, the burst recharge is provided by mass infall and mass transport across the disk by gravitational torques. In this respect, the bursts in our model are similar to GI-sustained MRI bursts in low-mass stars reported by \citet{Bae2014}. 
Our preliminary analysis, however, shows that the bursts in YIMSs are statistically longer in duration by about 1.0--1.5 orders of magnitude than in the corresponding MRI bursts for low-mass stars, given that the typical duration of the bursts for the LMSs being on the order of 100 years \citep[for LMSs, see further in][]{Vorobyov+2020b}. 
In addition, the accreted mass during individual bursts in YIMSs increases compared to LMSs, rising from a few tens of Jupiter masses to up to hundreds of Jupiter masses as mentioned earlier.
This is in agreement with simpler one-dimensional models of \citet{Elbakyan2021}. 
We also note that the photospheric luminosity rises with time for YIMSs \citep[see HR diagram in Fig 1 of][]{Vioque+2018}, which leads to luminosity bursts of smaller relative amplitude at later times, unlike the lower-mass cousins. The magnitude of the luminosity bursts are on average higher for young intermediate-mass stellar objects than for low-mass stars, reaching and exceeding $10^3~L_\odot$.
In addition to the above, 
the photospheric luminosity $L_{\rm *,ph}$ of a YIMS of final mass of 4.5~$\Msun$ (by 1~Myr) is about 10-500 times higher than that of a LMS of mass of 0.7~$\Msun$ as radius of such a YIMS is approximately 1.35–3.35 times larger than $R_{\odot}$ and $T_{\text{eff}}$ is about 1.5–2.9 times higher than $T_{\text{eff},\odot}$. 
Note that, in FEOSAD, $L_{\rm *,ph}$ and stellar radius $R_*$ are calculated using the stellar evolution tracks inferred with the STELLAR code of \cite{YorkeBodenheimer2008} \citep[see more details in][]{Vorobyov+2020}.

To clearly analyze the burst mechanism and demonstrate the link between the physical characteristics, we zoom in on a single episodic accretion outburst.  Figure \ref{fig:gasdust2DEvol} shows a detailed 2D spatial representation of the disk characteristics to understand how the burst is triggered within the inner region of PPD as a single accretion burst develops and decays. 
The outburst progresses in an outside-in fashion (see Figure \ref{fig:outsideinburst}).
The disk structure in the inner region is characterized by a spiral pattern in gas ($\Sigma_{\rm g}$) and rings with gaps in the grown dust ($\Sigma_{\rm d,gr}$). 
Such a layout in the gas and grown dust spatial morphology is indeed observed in several protoplanetary disks, spanning from tens to hundreds of au as observed in molecular line emissions \citep{Rosotti+2020,Huang+2023,Lee+2023}, dust scattered light \citep{Muto+2012,Benisty+2015,Benisty+2017,Muro-Arena+2020,Brown-Sevilla+2021}, and in millimeter dust continuum \citep{Andrews+2018,Long+2018,Cieza+2021}. It is likely caused by efficient dust drift towards local pressure maxima \citep{Birnstiel+2016,Vorobyov+2020b}. 
In our GI-dominated model, the pressure maximum is formed within $\sim$1.5-4~au of the inner disk because of the spatially varying rate of mass transport due to gravitational torques \citep{Vorobyov+2024}. The mass transport rate is strongest in the intermediate and outer disk regions where the $Q$ parameter is lowest and weakens in the inner several astronomical units, where the $Q$ parameter sharply rises (see Fig.~\ref{fig:2DQparDisk}). This effectively creates a bottleneck and a dead zone, similar in characteristics to the classic dead zone caused by the varying strength of MRI turbulence in the layered disk model \citep{Gammie1996}.

The higher concentration of dust in the massive dust ring makes the region optically thicker, 
which in turn slowly warms the gas in its vicinity and assists the process of MRI triggering (see Sec \ref{sec:viscosity_ion} for further details). 
The heating is provided by residual viscosity with $\alpha_{\rm visc} = 10^{-5}$ and adiabatic compression as the matter inflows along the spiral arms, creating a bottleneck effect until the thermal collisions begin to ionize the alkaline metals within the dead zone. Indeed, in the quiescent phase the ionization fraction is the lowest ($<10^{-12}$), but during the burst it increases several orders of magnitude. The midplane temperature $T_{\rm mp}$ rises from a few hundred Kelvin in the preburst stage to over a  few thousand Kelvin during the burst.   
Please refer to Figure \ref{fig:gasdust1DEvol} for the radial profiles of the azimuthally averaged disk quantities at these respective time instances as shown in Figure \ref{fig:gasdust2DEvol}. 
When the MRI is first triggered in the vicinity of the dust ring around 4-5~au, the $\alpha_{\rm visc}$ value also jumps to 10$^{-1}$, as suggested by the three-dimensional simulations of this phenomenon by \citet{Zhu2020}. The MRI front then propagates to both sides, but mostly inward, until it reaches the inner computational boundary at 0.5~au. At this time instance  $\dot{M}_{\ast}$ sharply increases, manifesting the onset of the accretion burst. The predominantly outward-in triggering of the burst in GI-supported disks was also reported in \citet{Bae2014}, its development is described in more details in their Sec~3, and it may have observational manifestations \citep{2023A&A...676A.124B,Nayakshin2024,2024MNRAS.529L.115G}. 

The process described above enhances rapid transport of the inner disk gas onto the star owing to sharply increased turbulent viscosity in an outside-in fashion. This phenomenon can be defined as MRI-triggered but GI-sustained bursts. The bursts lead to the complete destruction of the dust ring-like structures, which were formed earlier in the dead zones. The grown dust that was accumulated in the ring is first converted into small dust, the latter then transported onto the star along with gas because both are dynamically linked. The conversion of grown to small dust occurs because water ice mantles melt when the temperature rises during the burst and the dust aggregates break into monomers \citep{2017A&A...605L...2S}. We note that the recent analysis of the dust spectral index in V883~Ori, a young low-mass star in outburst, implied that icy dust mantles  can survive the heating events provided by the burst if water ice is a certain fraction of the solid matrix \citep{Houge2024}. This effect requires  further investigation, which goes beyond the scope of the present paper. Nevertheless, the dead zones are replenished and the dust rings start to form again soon after the burst ends (see Fig. \ref{fig:gasdust2DEvol}).

\subsection{Prospects of Streaming Instability}\label{sec:SI}
In this section, we focus on the dust growth, dust evolution, and
prospects of planetesimal formation in PPDs around YIMSs and present a comparison study on the development of  streaming instability (SI) in LMSs. 
Dust rings formed in our GI-controlled disk model may be favorable sites for planetesimal formation via SI. 
The SI is essentially a runaway situation that may occur when spatially varying gradients of gas pressure are present in the disk.
Once the dust-to-gas ratio increases due to the ensuing traffic jam of drifting dust particles, the back-reaction friction force from the dust onto the gas becomes strong enough to accelerate the gas to orbit at near-Keplerian speed. 
This reduces the headwind on the dust particles and slows down their inward drift \citep[see further in][]{Birnstiel+2016}. 
This altered flow then further enhances the dust concentration in those specific regions  
\citep{YoudinGoodman2005}. 
Without the direct modeling of SI, due to the limitations on higher resolution and modeling the vertical structure of gas and dust dynamics in our current work, we post-process the probable SI development in our model disk by taking into account the SI criteria obtained in focused multidimensional numerical simulations of gas-dust disks.

We investigate the favorable sites for the SI development in our model disk using the following criteria as presented in \citet[][hereafter Yang criterion]{Yang+2017} :
\begin{equation}\label{eq:SIYang1}
    {\rm log}\, \xi_{\rm d2g} \geq \, 0.10 \, ({\rm log\, St})^2 + \, 0.20 \, {\rm log \, St} - 1.76 \ ({\rm St} < 0.1),
\end{equation}
\vspace{-0.5cm}
\begin{equation}\label{eq:SIYang2}
    {\rm log}\, \xi_{\rm d2g} \geq \, 0.30 \, ({\rm log\, St})^2 + \, 0.59 \, {\rm log \, St} - 1.57 \ ({\rm St} > 0.1),
\end{equation}
where $\xi_{\rm d2g}= \Sigma_{\rm d,tot}/\Sigma_{\rm g}$ is the ratio of the total dust surface density to that of gas and $\mathrm{St}=\Omega_{\rm K} t_{\rm s}$ is the Stokes number, in which $\Omega_{\rm K}$ is the Keplerian orbital frequency and $t_{\rm s}$ is the dust stopping time (see Eq. \ref{eq:stokes}). 
By definition, $t_{\rm s}$ is the timescale over which the change in the relative velocity of a dust particle due to the drag force with the surrounding gas equals the value of velocity itself.
We further check the SI development in our model disk using the criterion \citep{YoudinGoodman2005} 
\begin{equation} 
\label{eq:rhoCriterion}
    \zeta=\frac{\rho_{\rm d, gr}}{\rho_{\rm g}} \geq 1.0  \, ,
\end{equation} 
i.e., the volume density of the grown dust in the disk midplane $\rho_{\rm d, gr}$ must be greater than or equal to that of the gas $\rho_{\rm g}$.
Here, the volume densities of gas and grown dust are calculated using the corresponding local vertical scale heights $H_{\rm g}$ and $H_{\rm d}$. 
The former is found from the assumption of the local vertical hydrostatic equilibrium, while the latter is derived using the following expression $H_{\rm d} = H_{\rm g} \sqrt{\alpha_{\rm visc}/(\alpha_{\rm visc}+\mathrm{St})}$ \citep{Birnstiel+2016}. 
In our numerical treatment, when $\mathrm{St} \ll \alpha_{\rm visc}$, such as during an outburst, $H_{\rm g}$ serves as the upper limit for $H_{\rm d}$ despite the respective jump in $\alpha_{\rm visc}$ going from an initial turbulent viscosity of about $10^{-4}$ to a maximum value of 0.1.  
Conversely, in the quiescent regime where $\mathrm{St} \gg \alpha_{\rm visc}$, 
$H_{\rm d}$ approximately drops as $H_{\rm g}\sqrt{\mathrm{\alpha_{\rm visc}/St}}$. 
We combine the two criteria (as shown in Eqs. \ref{eq:SIYang1}, \ref{eq:SIYang2}, and \ref{eq:rhoCriterion}) to employ a stricter constrain for finding the SI-prone regions (where both the criteria are fulfilled) in our model disk.

\begin{figure}
\epsscale{}
\plotone{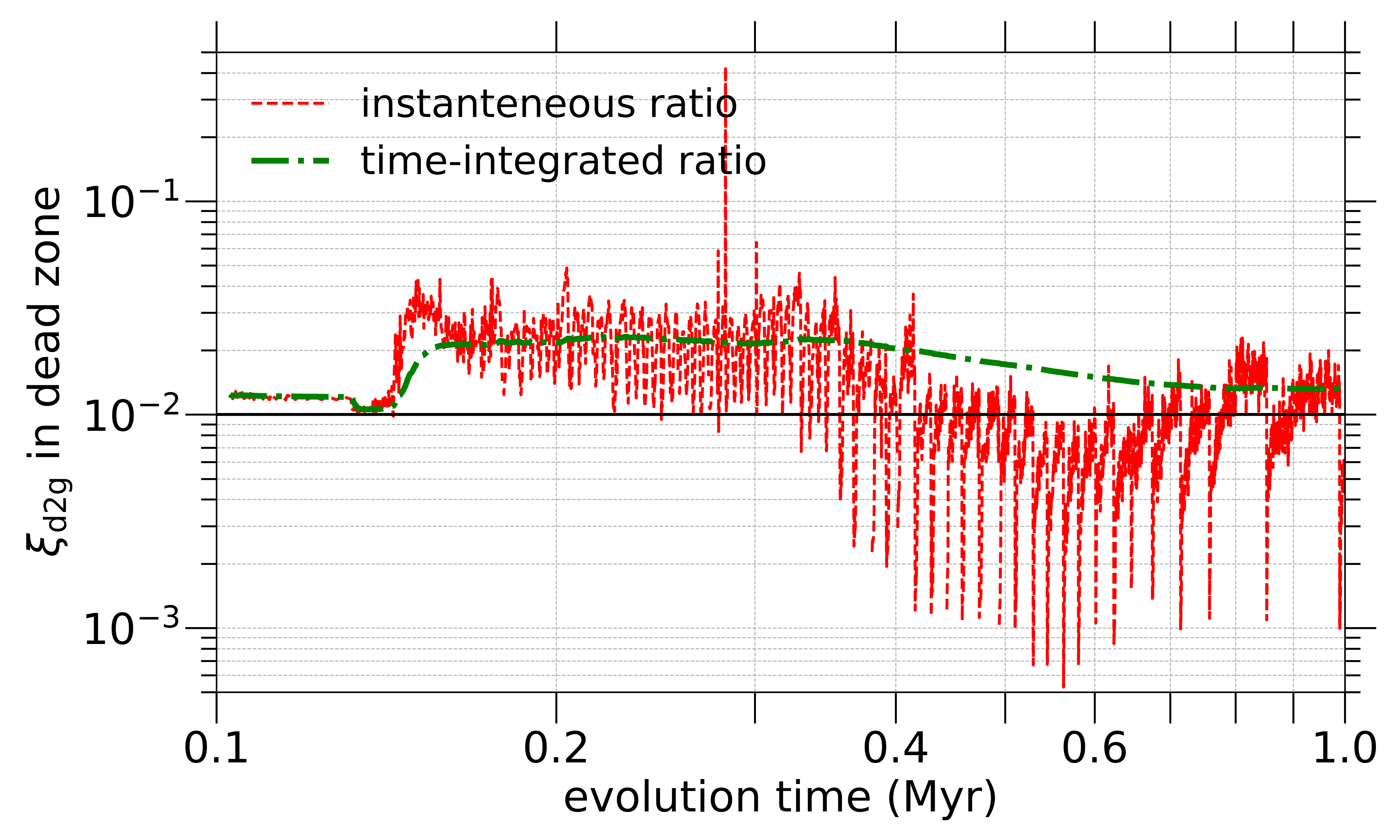}{(a)}
\plotone{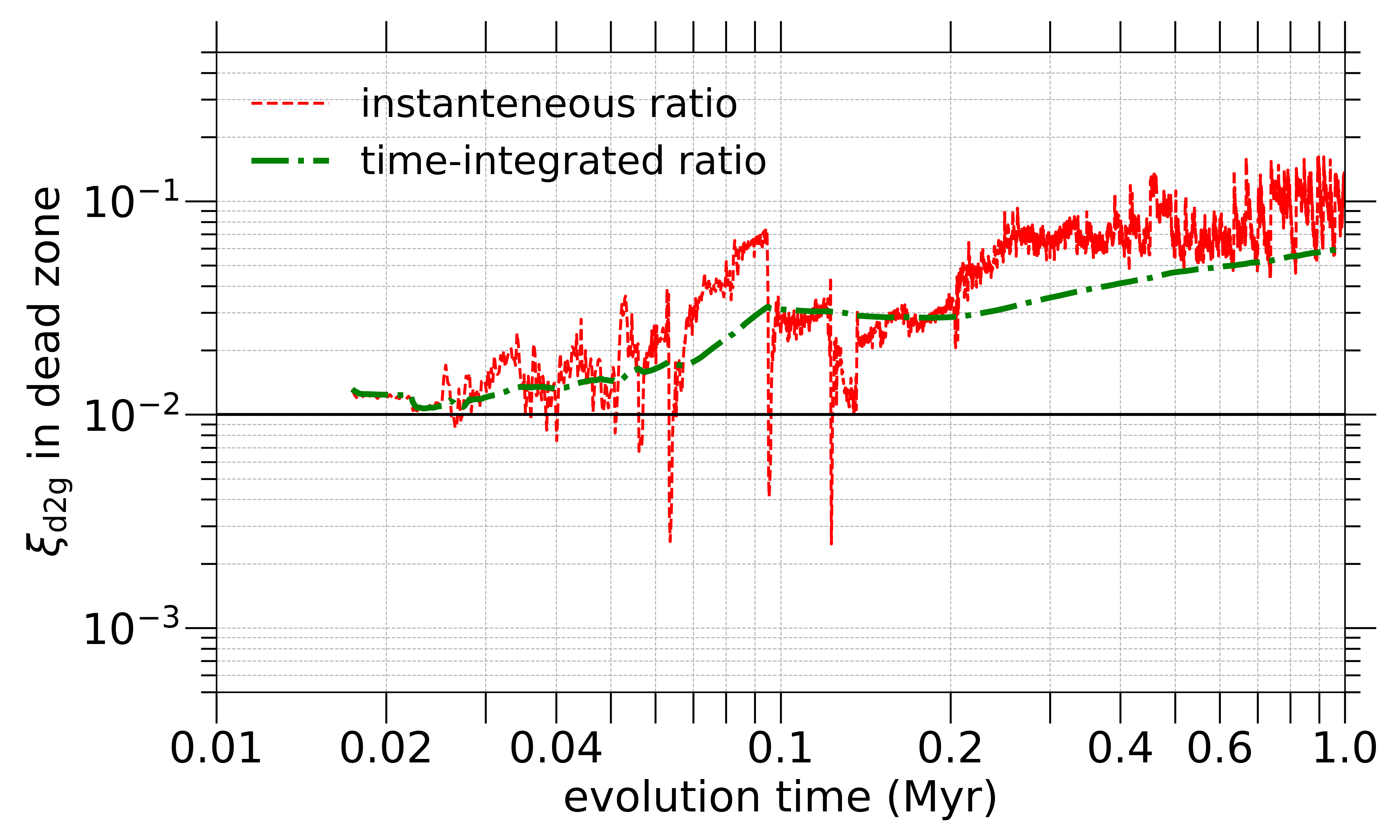}{(b)}
\caption{Temporal evolution of the instantaneous (red dashed line) and time integrated (green dash-dotted line) dust-to-gas mass ratios in the dead zone of the disk. The black solid horizontal line represents the initial canonical dust-to-gas mass ratio of $10^{-2}$. The top and bottom panel shows the cases for \simname{model-YIMS} and \simname{model-LMS}, respectively.}
\label{fig:massd2gDZ}
\end{figure}

The dust properties and the spacetime delineation of SI activity throughout the long-term evolution of the disk in \simname{model-YIMS} is presented in Figure \ref{fig:SISpacetime}, along with a direct comparison of the corresponding characteristics in \simname{model-LMS}. 
First panel of Figure \ref{fig:SISpacetime} shows the spacetime map of the total dust surface density for both models along with the enhanced dust rings. 
Indeed, for both the models, we examined the SI activity across all azimuths (refer to second panel of Fig.\ref{fig:SISpacetime}) along the dust rings as SI is a local phenomena. 
For the case of the disk around the YIMS,
the occurrence of SI is limited to the initial time period of disk evolution comprising several hundred thousand years and is spatially localized to the inner disk region within 10~au. 
In the late evolution after 0.35~Myr, the SI occurrence in YIMS is significantly suppressed as compared to that of around a LMS. 
Such a suppression of SI in PPDs around a YIMS is related to the systematic drop in the ratio of surface densities $\xi_{\rm d2g}$ (see further in Fig. \ref{fig:massd2gDZ}). Indeed, we checked the $\zeta$-criterion \citep[][refer to Eq.~\ref{eq:rhoCriterion}]{YoudinGoodman2005} and it is fulfilled in dust rings around the YIMS till the end of simulations. 
However, the Yang criterion (Eqs.~\ref{eq:SIYang1} and \ref{eq:SIYang2}) is not ubiquitously satisfied due to the insufficient $\xi_{\rm d2g}$ to exceed the threshold required to trigger SI activity as opposed to the case around the LMS. 
In our model, the Stokes number is calculated at the ${\rm a_{max}}$ (refer to Eq. \ref{eq:stokes}). 
For both models, within the dust rings, St increases above 1.0 due to a combined effect of larger ${\rm a_{max}}$ due to dust growth and  decrease in the midplane temperature of the dead zone.

The reason for the observed difference in the SI activity between LMSs and YIMSs can be manyfold. As noted in \citet{Pinilla2013} and \citet{Pinilla+2022}, the drift velocity of dust in a protoplanetary disk scales as $v_{\rm drift}\propto L_\ast^{1/4} / \sqrt{M_\ast} $. In \simname{model-LMS}, the stellar luminosity $L_\ast$ decreases  and the stellar mass $M_\ast$ saturates after the initial short period of burst activity. The resulting dust drift velocity declines as the disk evolves, thus assisting dust concentration in the dead zone. Conversely, the base stellar luminosity in \simname{model-YIMS} increases sharply with time, and in particular after 0.35~Myr, owing to the increasing photospheric luminosity as the YIMS makes the left turn on the H-R diagram. The stellar mass, however, grows slowly after 0.35~Myr. As a result, the dust drift velocity increases, which makes it more difficult for dust to concentrate in the dead zone and reduces $\xi_{\rm d2g}$. We note that dust settling in the dead zone may be efficient in both LMSs and YIMSs, so that the Youdin-Goodman criterion on the ratio of volume densities is fulfilled.

Furthermore, the evolution of the dead zone is affected by the burst activity, leading to the dead zone periodic destruction and rebuilding, a process that also reduces the rate of dust accumulation within the innermost disk regions (see Fig.~\ref{fig:2DSIcritburstv2} below). However, the adverse effect of the bursts on the SI activity is more severe for YIMSs than for LMSs because the former have many more bursts than the latter and the burst activity lasts longer (see Fig.~\ref{fig:1DEvolDiskV2}). Finally, the extent of the dead zone in \simname{model-YIMS} is greater than in \simname{model-LMS} because the disk of the former is more massive, making cosmic-ray ionization less efficient (see Fig.~\ref{fig:xIon}). Consequently, dust is spread
over a larger surface area of the dead zone in \simname{model-YIMS} and thus the ratio of dust to gas surface densities is systematically lower as opposed to \simname{model-LMS}.

Our analysis is confirmed by considering the instantaneous and time-integrated dust-to-gas mass ratios $\xi_{\rm d2g}$ in the dead zone for \simname{model-YIMS} and \simname{model-LMS} shown in Figure~\ref{fig:massd2gDZ}. The time-integrated $\xi_{\rm d2g}$ is found by firstly calculating the time-integrated masses of dust and gas and then taking their ratio. For the YIMS, both quantities decline after $t\approx0.35$~Myr, causing a sudden drop in the SI activity in these later times in agreement with the suggested increase in the radial dust drift velocity.  Sharp drops in the instantaneous ratio manifest the destructive effects of the bursts when the dead zone is destroyed and large amounts of dust that have accumulated are driven onto the star (see Fig.~\ref{fig:2DSIcritburstv2} below). For the LMS, both ratios generally increase with time, especially after the burst activity is subsided ($t>0.05$~Myr), promoting and sustaining the SI activity.

\begin{figure*}[htb!]
\centering
\begin{minipage}{\linewidth}
\centering
\includegraphics[width=\linewidth]{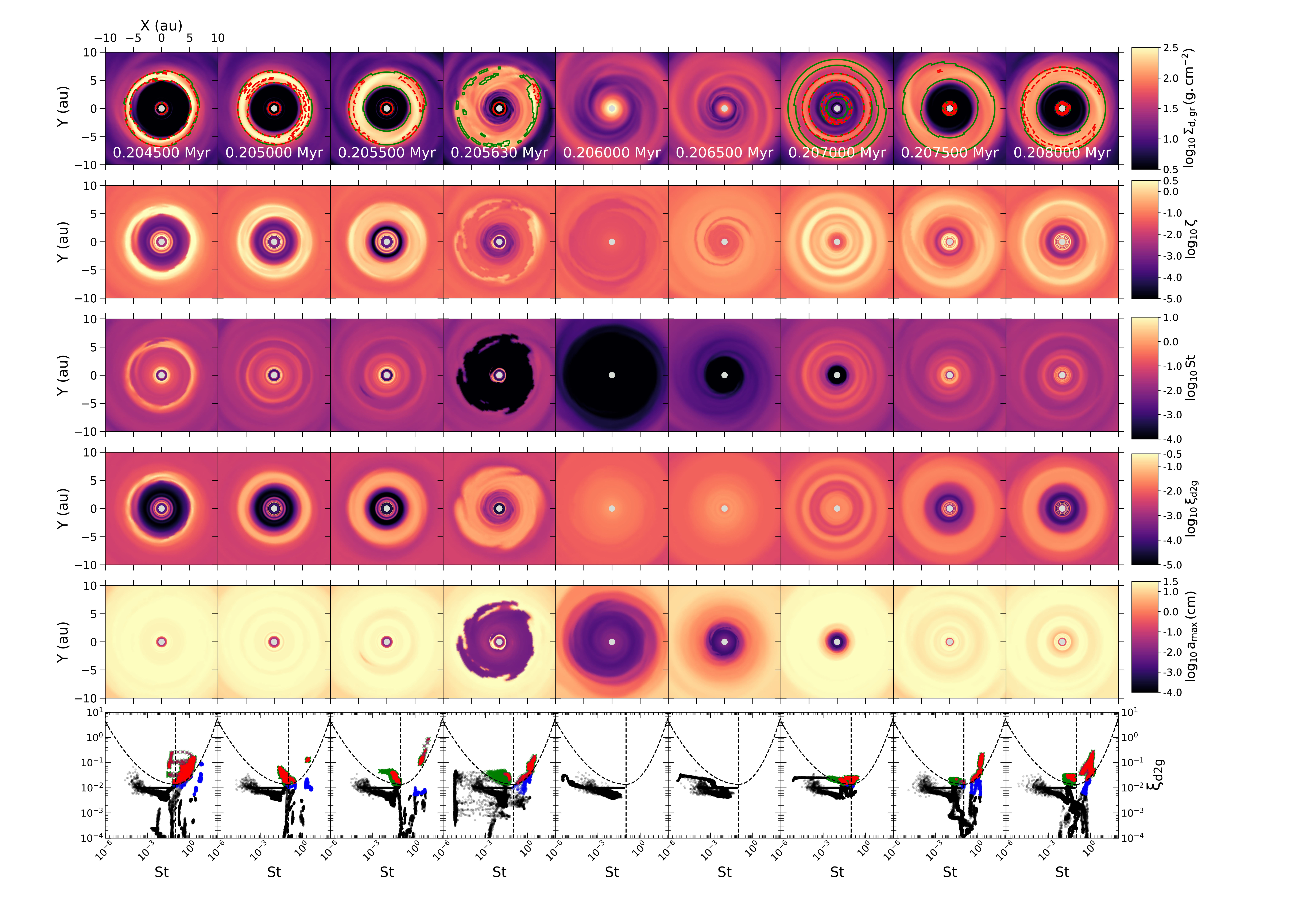}
\end{minipage}
\hfill
\begin{minipage}{\linewidth}
\vspace{-0.4cm}
\hspace{-1cm}
\centering        \includegraphics[width=0.8\linewidth,height=4cm]{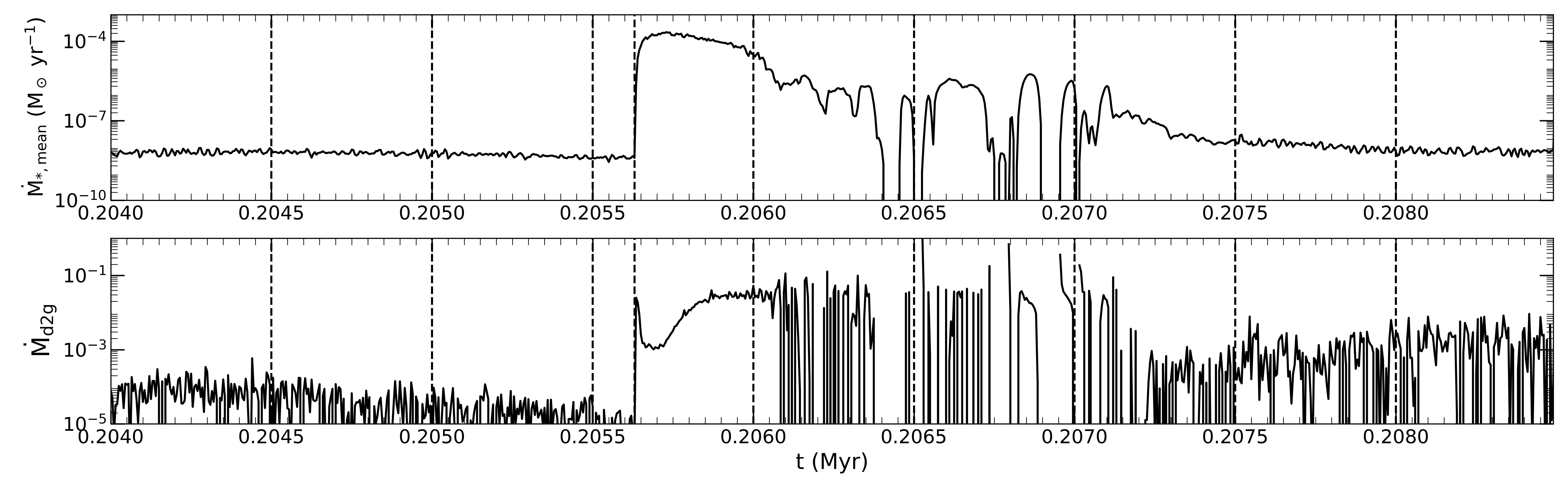}
\end{minipage}
\caption{SI development within the dust rings before and after an outside-in MRI outburst emerges as presented in 2D 
spatial maps of (top to bottom) grown dust surface density $\Sigma_{\rm d, gr}$ (in units of $\log_{10} \, {\rm g \, {cm}^{-2}}$), ratio of volume densities of grown dust to gas
$\zeta = \rho_{\rm d,gr}/\rho_{\rm g}$, Stokes number ${\rm St}$, total dust (grown+small) to gas mass ratio $\xi_{\rm d2g} =\Sigma_{\rm d,tot}/\Sigma_{\rm g}$, given $\Sigma_{\rm d,tot}= \Sigma_{\rm d,gr} +\Sigma_{\rm d,s}$, maximum size of dust grains $a_{\rm max}$ (in units of $\log_{10} \, {\rm cm}$), and the phase space of $\xi_{\rm d2g}$ versus ${\rm St}$ showcasing SI development over a region of $20 \times 20 \, {\rm au}^2$ as obtained from \simname{model-YIMS}. 
In the 2D map of $\sigmadustg$, the green contour refers to the SI-prone according to \cite{Yang+2017}-criterion only. Whereas, the contours in red belongs to an SI-prone zone where the \cite{Yang+2017}-criterion (Eqs.~\ref{eq:SIYang1} and \ref{eq:SIYang2}) as well as the \cite{YoudinGoodman2005}-criterion (Eqs.~\ref{eq:rhoCriterion}) are satisfied simultaneously.
In the phase space diagram, the green and blue scattered dots represent the distribution of region prone to SI following the Yang-criterion only and Youdin-Goodman criterion only, respectively. The black-dashed curve marks the critical condition for the onset of SI as per equation \ref{eq:SIYang1} (for ${\rm St} < 0.1$) and \ref{eq:SIYang2} (for ${\rm St} > 0.1$). The black-dashed vertical line is a dividing line set at  ${\rm St} = 0.1$. 
The overlaid red dots correspond to the SI-prone region where both the criteria are fulfilled. 
The bottom most panel shows the temporal evolution of mass accretion rate onto the star and dust-to-gas-mass accretion rate onto the star at the sink-disk interface for \simname{model-YIMS} as an outside-in MRI outburst progresses. The vertical dashed lines mark the time instances corresponding to the above 2D intensity maps.} 
\label{fig:2DSIcritburstv2}
\end{figure*}

Figure \ref{fig:2DSIcritburstv2} portrays the time evolution of the SI development within the inner region of the PPD from the preburst phase until the postburst phase in \simname{model-YIMS}. 
From the spatial distribution of $\xi_{\rm d2g}$, $\mathrm{St}$, and $\zeta$ we determine the SI-prone zones within the inner disk, shown by the contour lines. In the preburst stage, the SI-prone regions are mostly localized in the vicinity of the dust ring. However, as the burst is triggered, the physical conditions become unfavourable for development of the SI. In particular, the Stokes number drops significantly because most of the grown dust is converted into monomers with a size of several $\upmu$m. The ratio $\zeta$ consequently drops as well. The emptying of gas and dust from the MRI-active region during the burst completes the SI termination. In the postburst stage, the dust rings reform, dust regrows to its original size, and the SI-prone regions appear again. The bottom panel in Figure~\ref{fig:2DSIcritburstv2} demonstrates that protostellar mass accretion is enriched with dust, reinforcing the scenario of dust ring destruction and accretion onto the star.

\begin{figure*}
\vspace*{-.4cm}
\gridline{\fig{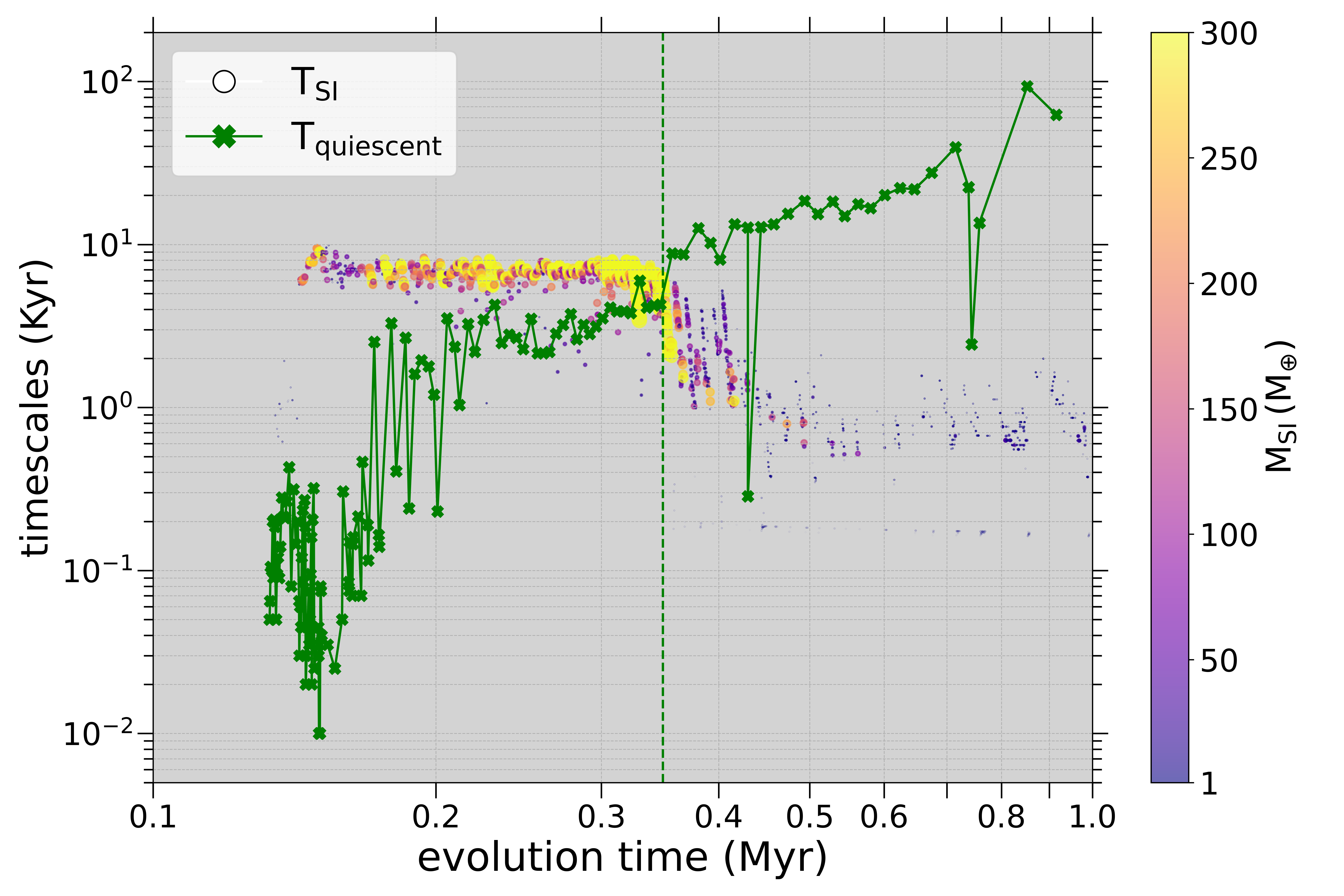}{0.5\textwidth}{(a)}
\fig{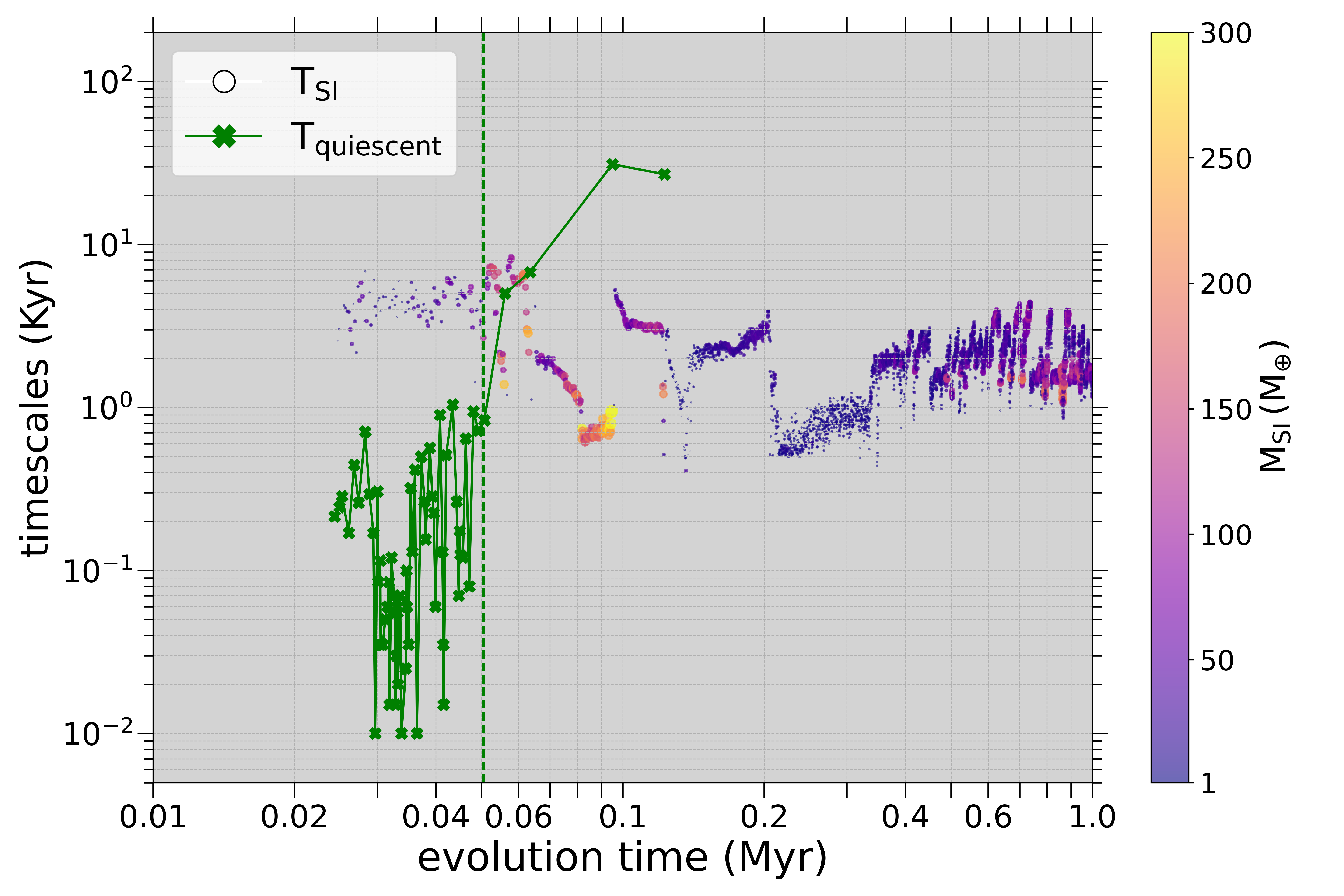}{0.5\textwidth}{(b)}}
\vspace*{-.35cm}
\caption{  
Comparison between the quiescent period ${T_{\rm quiescent}}$ (i.e, time between significant outbursts) to the approximate growth timescale of streaming instability ${T_{\rm SI}}$ 
overlaid with the grown dust mass (in the units of $\mearth$) available for SI-development in the disk as fulfilled by the $\xi_{\rm d2g}$-criterion as well as the $\zeta$-criterion (as shown in Eqs. \ref{eq:SIYang1}, \ref{eq:SIYang2}, and \ref{eq:rhoCriterion}). 
The left (a) and right (b) panel shows the cases for \simname{model-YIMS} and \simname{model-LMS}, respectively.}
\label{fig:TSI_TQuiescent}
\end{figure*}

Frequent bursts, as observed in the early evolution of our model, may impede the feasibility of planetesimal formation via the SI even though the inner disk is formally prone to develop the SI.  
This can occur if the time between bursts is shorter than the characteristic time of planetesimal formation via the streaming instability. 
Figure \ref{fig:TSI_TQuiescent}  shows a quantitative comparison between the quiescent period, ${T_{\rm quiescent}}$, and the approximate mass-weighted growth timescale of the streaming instability, ${T_{\rm SI}}$ alongside the estimates of the instantaneous $M_{\rm SI}$ for \simname{model-YIMS} and \simname{model-LMS}. 
Here, ${T_{\rm quiescent}}$ presents the duration between two successive FU-Ori outbursts of significant magnitude. 
A threshold of a $40\%-60\%$ increase in the disk midplane temperature relative to its quiescent period value, along with continuously surpassing the ionization temperature of alkaline metals ($\sim \,1000\,{\rm K}$), are chosen as criteria to define the bursts in this study. 
Dust-gas coupled hydrodynamic simulations in the
shearing-box approximation suggest that the timescale of development
for the streaming instability is typically of the order of 100-1000 times the local orbital period \citep{Simon+2016,Yang+2017}. 
In our work, we calculate mass weighted SI growth timescale as follows 
\begin{equation}
    T_{\rm SI} = \frac{\sum_{i} (M_{\rm SI}^{i} * t_{\rm SI}^{i})}{\sum_{i} M_{\rm SI}^{i}}
\label{eq:TSI-Mweighted}
\end{equation}
where $M_{\rm SI}^{i}= \Sigma_{d,gr}^{i}{\rm (SI)}\,* \Delta A^{i}$, $t_{\rm SI}^{i} = 2\pi F/\Omega_{K}^{i}$, $F$ = 500, $\Omega_{K}^{i}$ and $\Delta A^{i}$ represent the (Keplerian) orbital speed and the area for that respective cell in the computational grid, respectively. 
In our calculation, $M_{\rm SI}$ is estimated based on the grown dust mass available for the proposed SI activity as fulfilled by the $\xi_{\rm d2g}$-criterion as well as the $\zeta$-criterion (as shown in Eqs. \ref{eq:SIYang1}, \ref{eq:SIYang2}, and \ref{eq:rhoCriterion}).

We see that in the early evolution phase, the quiescent period is significantly shorter or only marginally comparable to the SI-growth timescale, which gives the SI almost little to no chance to grow within the dead zone of the inner disk in this early evolution time period. 
The vertical green dashed line in Figure \ref{fig:TSI_TQuiescent} represents the time instant when the quiescent period becomes longer than the SI timescale.
The time when $T_{\rm quiescent}$ becomes longer than $T_{\rm SI}$ and frequent bursts cannot anymore prohibit the SI development is $t \approx 0.35$~Myr for \simname{model-YIMS}.
Despite the longer duration of the quiescent period after $\approx 0.35$ Myr in this model, planetesimal formation via SI is inefficient at these later times due to fast dust drift velocities and insufficient $\xi_{\rm d2g}$ as discussed earlier. 
For \simname{model-LMS}, however, planetesimal formation can commence as early as $t \approx 0.05$~Myr and continue throughout the entire evolution period.
This result supports the idea that planetesimal formation via SI may be difficult in young intermediate-mass stars, which is consistent with the observed dearth of planets around stars with $M_{\ast} > 3.0$~${\rm M}_\odot$.  
A similar comparison of these two timescales was investigated by \citet{Kadam+2022} for the low-mass protostars and showed that ${T_{\rm quiescent}}$ exceeded ${T_{\rm SI}}$ already in the Class I stage ($\sim 0.1$~Myr). 
Moreover, disk conditions continue to be favourable for SI during the first 1.0~Myr of disk evolution, thus promoting planet formation around low-mass stars.

\section{Discussions} \label{sec:discussion}

Our results show the long-term history of episodic accretion in PPDs around YIMSs and prospects of the first-generation planetesimal formation, especially without the requirement for a pre-existing planet to carve out pressure traps. 
A comparative study is also conducted on SI activity in the PPDs around LMSs.
The unique advantage of using FEOSAD code is that it can self-consistently globally study the formation and evolution of a PPD starting from the collapse phase of a magnetized prestellar cloud core without considering the typical assumption
of a minimum mass solar nebula \citep[MMSN;][]{Adams+2010} as an initial condition, thus able to include increasingly complex physics. 
Our current framework also considers the co-evolution of dynamically coupled dust-gas, including its back-reaction on gas and dust growth. 
As a result, it provides insights that are difficult to attain through 1D simulations using simplified viscous disk models \citep{Pringle1981} or 3D simulations that are restricted to a short duration or cover only a partial region of the PPD disk, such as when using a shearing box. 

In this work, we focused on the formation of GI-induced dust rings within the innermost few au of the PPD around a YIMS. 
While being initially susceptible to the SI, these rings appear to be unfriendly environments to planetesimal formation. First, continual MRI ignition destroys inner dust rings by accreting matter onto the star from the outside-in. 
This temporarily shuts down the SI development. 
Second, most of the SI activity is limited to the initial 0.35~Myr of disk evolution, but during this time period the duration of the quiescent phase between the bursts is shorter than or comparable to the characteristic time of planetesimal formation via SI (refer to Fig \ref{fig:TSI_TQuiescent}a). This may also reduce the efficiency of planetesimal formation in the protoplanetary disks of YIMSs. 
In the later evolution, an increase in the dust drift velocity makes dust concentration in the rings less efficient, thus suppressing the SI activity.
Finally, some fraction of dust may be sublimated during the bursts as our preliminary estimates suggest (see Sect.~\ref{sec:DustEvapApp}).
These factors when taken together can explain the dearth of planets around stars with masses above $3 \, \Msun$. 
Whereas, planetesimal formation in disks around LMSs does not suffer from these adverse effects to the extent sufficient to shut down the process, a conclusion also reached in \citet{Kadam+2022}. 
Our calculation shows that for LMSs planetesimal formation can proceed after $t \approx 0.05$~Myr onwards (refer to Fig \ref{fig:TSI_TQuiescent}b).

According to our preliminary analysis, though SI activity is significantly suppressed in YIMSs, however, sporadically may occur within 3--4~au (refer to second row of Fig. \ref{fig:SISpacetime}).
At such distances from the central YIMS, the paucity of planets around stars more massive than about $3 \, \Msun$ may be a selection effect since the radial velocity effect becomes smaller as the mass of the host star increases, and transits are also more difficult to measure against brighter background. 
However, the observed sharp drop at about $3 \, \Msun$, rather than a smooth transition from detection to no-detection, can potentially indicate a physical effect and our study proposes one 
possible physical solution.

Our simulations are performed for a moderately supercritical massive prestellar core with the evolution of magnetic field under the approximation of flux-freezing. 
Because of the high computational cost, 
the nonideal MHD effects (ambipolar diffusion and Ohmic dissipation) have been excluded, but they might be important in the optically thick and cold environments of PPDs \citep[for the low-mass stars, see][]{Turner+2014}. 
The diffusion and dissipation of the magnetic field in the innermost regions of a PPD can alter the radial extent of the dead zone by reducing $\Sigma_{\rm MRI}$ (gas column density of MRI-active region) and MRI turbulence, which in turn may affect the frequency of MRI outbursts.

In this work, we estimated the mass-weighted SI-timescale ($T_{\rm SI}$) to require 500 times the orbital period (refer to Eq.~\ref{eq:TSI-Mweighted}). The choice of such a factor of 500 is often obtained based on the analysis from the shearing-box simulations \citep{Yang+2017} that treat the SI activity as a standalone process without considering the global PPD evolution. 
However, the episodic burst activity 
may alter the SI timescale. Indeed, a burst destroys the SI-prone zones in the inner disk and additional time may be required to rebuild the sufficient enhancement in $\xi_{\rm d2g}$ and $\zeta$, and for dust to regrow to the proper size and Stokes number \citep{Vorobyov2022},  within the dead zone such that it can further trigger SI activity.

In our current study, FEOSAD calculates the optical depth as $\Sigma_{\rm d,tot} \kappa_{d}$, where $\Sigma_{\rm d,tot}$ is the column density of dust and $\kappa_{d}$ is the mean opacity taken from \cite{Semenov+2003} and scaled per unit mass of dust. Although dust sublimation is not explicitly realized in the model, Semenov et al. opacities take into account the effects of dust sublimation, resulting in a sharp drop in opacities above $\sim 1000 \, {\rm K}$. We also note that Semenov et al. opacities do not depend on dust growth and more accurate computations on dust-growth dependant opacities are required to confirm the effect. 
In addition to this, the possible enhancement in the X-ray luminosity of the accreting YIMS during an outburst by several orders of magnitude \citep{2014A&A...570L..11L} may affect the structure of the innermost dead zone and following outburst characteristics.

\section{Conclusions} \label{sec:conclusions}
In this study, we present the results of numerical simulations using dynamically coupled gas-dust global MHD disk models, which account for disk self-gravity and examine the long-term evolution of a PPD around a young intermediate-mass star (YIMS) since the collapse of the prestellar cloud core. 
Gravitationally unstable disk models with globally suppressed but episodically triggered MRI in the inner disk are considered. 
The focus of the study is concentrated on the ensuing episodic MRI-triggered accretion bursts and prospects of planet formation via the SI in the dust rings within the PPD.  A comparison study with the SI activity in disks of low-mass stars (LMSs) was also carried out.
Here, we summarize the key findings of this study as follows. 
\begin{enumerate}

    \item Accretion bursts are ubiquitous in YIMSs, in particular, in the embedded stages of their evolution. These MRI-triggered but GI-sustained bursts are typically $\sim 10$ times longer than in their low-mass star counterparts. 
    
    \item Massive concentrations of gas and dust form within the inner disk of a YIMS owing to the radially varying efficiency of mass  transport via GI, with dust often shaped into rings. These rings, while initially susceptible to streaming instability, tend to be destroyed over time due to frequent accretion bursts and dust sublimation. The time between the bursts is initially shorter than the timescale of SI growth, further reducing the efficiency of planetesimal formation in the quiescent interburst periods.

    \item  
    In the later disk evolution of a YIMS, when the burst activity subsides and the quiescent periods become longer than the SI growth timescale, an increased dust drift velocity owing to the growing stellar luminosity makes dust concentration in the inner disk less efficient. A larger extent of the dead zone in disks of YIMS compared to that in around LMSs also leads to smaller ratios of dust-to-gas surface densities, thus significantly reducing the SI activity in the disk of YIMS. 

    \item
    Conditions for the SI development in disks around LMSs appear to be less affected by luminosity bursts and the variations in the dust drift velocity, implying efficient planetesimal formation throughout most of the initial 1.0~Myr of disk evolution.
 \end{enumerate}   
 
Our results suggest that the likelihood for planetesimal formation  around young intermediate-mass stars is significantly reduced, in agreement with the observed dearth of planets around these objects.

\section*{Acknowledgments}
We thank Kundan Kadam for helpful discussions. 
We are thankful to the anonymous referee for constructive
comments and suggestions that helped to improve the manuscript.
The numerical simulations were performed with High-Performance Computation (HPC) support provided by the Digital Research Alliance of Canada (\href{https://alliancecan.ca/en}{alliancecan.ca}) and the regional partner organizations \href{https://www.computeontario.ca/}{Compute Ontario} and \href{https://www.calculquebec.ca/}{Calcul Québec} on the HPC clusters \href{https://docs.alliancecan.ca/wiki/Graham}{Graham},  \href{https://docs.alliancecan.ca/wiki/Cedar}{Cedar},  
and \href{https://docs.alliancecan.ca/wiki/Narval}{Narval}. 
ID acknowledges grant support for Theory from the Institute of Astronomy and Astrophysics, Academia Sinica (ASIAA), and the National Science and Technology Council (NSTC) in Taiwan through grant 113-2112-M-001-008. 
The work of E.I.V. from UrFU for Sects. 2
and 3 was supported by the Russian Science Foundation, project no. 23-12-00258.
SB was supported by a Discovery Grant from NSERC and acknowledges the hospitality of ASIAA and the University of Vienna at different stages of this research.
We thank Oliver Herbort for providing the data from the ggChem model. 
ID also acknowledges the use of 
the TARA compute server
at the Department of Physics \& Astronomy of the University of Western Ontario, Canada.

\appendix

\section{Numerical Methods} \label{sec:appmethods}
\subsection{Numerical Equations for gaseous component} \label{sec:gas}
The equations of mass continuity, momentum conservation, and energy transport are solved in the thin-disk approximation, which can be written as follows
\begin{equation}
\label{eq:cont}
\frac{\partial \Sigma_{\rm g}}{\partial t}  + \vec{\nabla}_{\rm p}  \cdot \left(\Sigma_{\rm g} \vec{v}_{\rm p} \right) = 0,  
\end{equation}
\begin{eqnarray}
\label{eq:mom}
\frac{\partial}{\partial t} \left( \Sigma_{\rm g} \vec{v}_{\rm p} \right) + \left[ \vec{\nabla} \cdot \left( \Sigma_{\rm g} \vec{v}_{\rm p} \otimes \vec{v}_{\rm p} \right) \right]_{\rm p}  =   - \vec{\nabla}_{\rm p} {\cal P}  + \Sigma_{\rm g} \, \left( \vec{g}_{\rm p} +\vec{g}_\ast \right) 
+ \left(\vec{\nabla} \cdot \mathbf{\Pi} \right)_{\rm p}  - \Sigma_{\rm d,gr} \vec{f}_{\rm p} +  {B_z {\vec B}_p^+ \over 2 \pi} - H_{\rm g}\, \vec{\nabla}_{\rm p} \left({B_z^2 \over 4 \pi}\right), ~ ~ ~
\end{eqnarray}
\begin{equation}
\frac{\partial e}{\partial t} +\vec{\nabla}_{\rm p} \cdot \left( e \vec{v}_{\rm p} \right) = -{\cal P}
(\vec{\nabla}_{\rm p} \cdot \vec{v}_{\rm p}) -\Lambda +\Gamma + 
\left(\vec{\nabla} \vec{v}\right)_{pp'}:\Pi_{pp'}, 
\label{eq:energy}
\end{equation}
respectively, where the subscripts $p$ and $p'$
refer to the planar components
$(r, \phi)$ in polar coordinates, 
$\Sigma_{\rm g}$ is the gas surface density, $\vec{v}_{\rm p} = v_r \hat{r} + v_{\phi} \hat{\phi}$ is the gas velocity in the disk plane, $\nabla_{\rm p} = \hat{r} \partial/\partial r + \hat{\phi} r^{-1} \partial/\partial \phi$ is the gradient along the planar coordinates of the disk, 
${\cal P}$ is the vertically integrated gas pressure, 
The ideal gas equation of state is used to calculate the vertically integrated gas pressure, ${\cal P}=(\gamma-1) e$ with $\gamma=7/5$. 
The gravitational acceleration in the disk plane $\vec{g}_{\rm p} = g_r \hat{r}+ g_{\phi} \hat{\phi}$, takes into account self-gravity of disk by solving for the disk gravitational potential
using the Poisson integral \citep[see details in ][]{BT2008, Vorobyov+2024}.
The term $\vec{g}_{\ast}$ is the gravitational acceleration due to the central protostar, which only has a radial component.
The viscous stress $\mathbf{\Pi}$ tensor is defined as $\mathbf{\Pi}=  2\Sigma_{\rm g} \nu \left({\nabla} \vec{v} -(\nabla \cdot \vec{v})\vec{1}/3 \right)$, where $\nu$ is the kinematic viscosity, $\vec{1}$ is the unit tensor, and $\vec{\nabla}v$ is the symmetrized velocity gradient tensor (see \S 2 and Appendix B of \citet{VorobyovBasu2009} and also in \citet{VorobyovBasu2010}).
The term $\vec{f}_{\rm p}$ is the drag force per unit mass between dust and gas, and $\Sigma_{\rm d,gr}$ is the surface density of grown dust (see details in Appendix \ref{sec:dust}). 
The code is written in the thin-disk limit, complemented by a calculation of the gas vertical scale height ($H_{\rm g}$) using an assumption of local hydrostatic equilibrium in the gravitational field of disk and star, as described in \cite{VorobyovBasu2009}. 
The resulting model has a flared structure (because
the disk vertical scale height increases with radius), which guarantees that both the disk and envelope receive a fraction of
the irradiation energy from the central protostar. 
Here, $B_z$ is the vertically constant but radially and azimuthally varying $z$-component of the magnetic field within the disk thickness. 
The planar components of the magnetic field are ${{\vec B}}_p^+ = {B}_r^+ \hat{r} + {B}_{\phi}^+ \hat{\phi}$ where the `$+$' corresponds to the component at the top surface of the disk. 
Furthermore, 
$e$ is the internal energy per unit surface area, $\Gamma$ and $\Lambda$ represent the cooling and heating rates as discussed later in Eq. (\ref{eq:cooling}) and (\ref{eq:heating}), respectively.

Coming to the magnetic field physics \citep[see also \S 2.3 in][]{Vorobyov+2020}, 
the planar component of the magnetic field at the top surface of the disk is denoted by ${\vec B}_p^+$ and the midplane symmetry is assumed, such that ${{\vec B}}_p^-=- {\vec{B}}_p^+$. 
In Equation (\ref{eq:mom}) the last two terms on the right-hand side are the Lorentz force (including the magnetic tension term) and the vertically integrated magnetic pressure gradient. 
The magnetic tension term arises
formally due to the Maxwell stress tensor, that can be intuitively understood as the interaction of an electric current at the disk surface 
\citep[see also Figure 1 of][]{DasBasu2021}. 
The discontinuity in tangential field component gives rise to surface current while there is no current within the disk. 
In the adopted thin-disk approximation, the magnetic field
and gas velocity have the following nonzero components in
the disk: $\vec{B}= (0, 0, B_z)$ and $\vec{v} = (v_r, v_{\phi}, 0)$, respectively. 
The vertical component of magnetic field is calculated by explicitly solving the induction equation in the ideal MHD regime:  
\begin{eqnarray}
\frac{\partial B_z}{\partial t} = -\frac{1}{r} \left( \frac{\partial}{\partial r}\left(r v_r B_z\right) + \frac{\partial}{\partial \varphi}\left(v_{\varphi}B_z\right)\right),
\label{eq:Bz}
\end{eqnarray}
wherein the advection of $B_z$ is considered. 
The diffusive effects of Ohmic dissipation and ambipolar diffusion are neglected due to high computational cost. The total magnetic field can be written as the gradient of a scalar magnetic potential $\Phi_{\rm M}$ and the planar component of magnetic field ($\vec{B}_p^+$) is calculated by solving the Poisson integrals \citep[see details][]{Vorobyov+2020} with the source term of $(B_z - B_0)/(2 \pi)$, where $B_0$ is the constant background magnetic field with value of $10^{-5} \, {\rm G}$.

The heating and cooling rates $\Gamma$ and $\Lambda$, respectively are based on the analytical calculations of the radiation transfer in the vertical direction \citep{DongEtal2016}. 
The equation for the cooling rate is
\begin{equation}
    \Lambda = \frac{8 \tau_{\rm P} \sigma_{\rm SB} T_{\rm mp}^4}{1+2 \tau_{\rm P}+1.5 \tau_{\rm R} \tau_{\rm P}} \ ,
    \label{eq:cooling}
\end{equation}
where $T_{\rm mp} = {\cal P} \mu/{\cal R} \Sigma_{\rm g}$ is the midplane temperature, $\mu=2.33$ is the mean molecular weight, $\cal{R}$ is the universal gas constant, and $\sigma_{\rm SB}$ is the Stefan-Boltzmann constant. Here, $\tau_{\rm P}= 0.5 \Sigma_{\rm d, tot} \kappa_{\rm P}$ and $\tau_{\rm R}= 0.5 \Sigma_{\rm d, tot} \kappa_{\rm R}$ represent the Planck and Rosseland optical depths to the disk midplane, where $\kappa_{\rm P}$ and $\kappa_{\rm R}$ are the Planck and Rosseland mean opacities taken from \cite{Semenov+2003} and scaled to the unit mass of dust.  The optical depths in the calculations are proportional to the total dust surface density ($\Sigma_{\rm d,tot}$).

The heating function takes into account the stellar irradiation at the disk surface as well as background blackbody irradiation. The heating function per unit surface area of the disk is 
\begin{equation}
    \Gamma = \frac{8 \tau_{\rm P} \sigma_{\rm SB} T_{\rm irr}^4}{1+2\tau_{\rm P}+1.5\tau_{\rm R} \tau_{\rm P}} \ ,
    \label{eq:heating}
\end{equation}
where $T_{\rm irr}$ is the irradiation temperature at the disk surface and
\begin{equation}
    T_{\rm irr}^4 = T_{\rm bg}^4 + \frac{F_{\rm irr}(r)}{\sigma_{\rm SB}} \ .
\end{equation}
Here, $T_{\rm bg}$ represents the temperature of the background blackbody irradiation, $F_{\rm irr}(r) = L_* \cos \gamma_{\rm irr}/(4\pi r^2)$ is the radiation flux (i.e., energy per unit
surface area per unit time) absorbed by the disk surface at a radial distance $r$ from the central star having a stellar luminosity $L_*$. The $L_*$ is a sum of the accretion luminosity $L_{\rm *,accr} = 0.5 \, GM_* \dot{M}/R_*$ 
arising from the gravitational energy of accreted gas plus the photospheric luminosity $L_{*, {\rm ph}}$ due to gravitational compression and
deuterium burning in the stellar interior. Here, $\dot{M}$, $M_*$, and $R_*$ are the mass accretion rate onto the star, the stellar mass, and the radius of the star, respectively. 
Furthermore, $\gamma_{\rm irr}$ is the incidence angle of radiation that arrives at the disk surface w.r.t. the normal at a radial distance $r$ \citep[see Eq. (10) of \S 2.1 of][]{Vorobyov+2020}.
The resulting model has a flared structure, wherein the disk vertical scale height increases with radius.
Both the disk and the envelope receive a fraction of the irradiation energy from the central protostar.
The adopted opacities in the numerical model of FEOSAD do not take dust growth into account.

\subsection{Numerical Equations for Dust Components} \label{sec:dust}
The dust in the numerical model of FEOSAD is based on two components, small dust and grown dust \citep[see][for more details]{Vorobyov+2018, Vorobyov+2020}. 
In our numerical model, small dust has a grain size of $a_{\rm min}<a<a_\ast$, and grown dust corresponds to a size of $a_\ast \le a<a_{\rm max}$, where $a_{\rm min}=5\times 10^{-3}$~$\upmu$m, $a_\ast=1.0$~$\upmu$m. Here, $a_{\rm max}$ is a dynamically varying maximum radius of the dust grains, which depends on the efficiency of radial dust drift and the rate of dust growth. 
Small dust is dynamically coupled with the gas as it is sub-micron dust grains by definition.
Whereas, grown dust dynamics is regulated by friction with gas and the total gravitational potential of the star, gas, and dust components. 
All dust grains combined are assumed to have a material density of $\rho_{{\rm s}}=3.0\,{\rm g~cm}^{-3}$.
The equations of continuity and momentum conservation for small and grown dust components in the pressure-free limit are as follows:
\begin{equation}
\label{eq:contDsmall}
\frac{{\partial \Sigma_{\rm d,sm} }}{{\partial t}}  + \vec{\nabla}_{\rm p} \cdot 
\left( \Sigma_{\rm d,sm} \, \vec{v}_{\rm p} \right) = - S(a_{\rm max}),  
\end{equation}
\begin{equation}
\label{eq:contDlarge}
\frac{{\partial \Sigma_{\rm d,gr} }}{{\partial t}}  + \vec{\nabla}_{\rm p}  \cdot 
\left( \Sigma_{\rm d,gr} \, \vec{u}_{\rm p} \right) = S(a_{\rm max}),  
\end{equation}
\begin{eqnarray}
\label{eq:momDlarge}
\frac{\partial}{\partial t} \left( \Sigma_{\rm d,gr} \, \vec{u}_{\rm p} \right) +  \left[\vec{\nabla} \cdot \left( \Sigma_{\rm
d,gr} \, \vec{u}_{\rm p} \otimes \vec{u}_{\rm p} \right)\right]_{\rm p} =  
  \Sigma_{\rm d,gr}  \left( \vec{g}_{\rm p}
  + \vec{g}_\ast \right) \,
 + \Sigma_{\rm d,gr} \vec{f}_{\rm p} + S(a_{\rm max})  \,\vec{v}_{\rm p},
\end{eqnarray}
where $\Sigma_{\rm d,sm}$ and $\Sigma_{\rm d,gr}$ are defined as the surface densities of small and grown dust, respectively. The term $\vec{u}_{\rm p}$ describes the planar components of the grown dust velocity, and $S(a_{\rm max})$ is the rate of conversion from small to grown dust per unit surface area and it is a function of the fragmentation size of the dust.  The rate of small to grown dust conversion $S(a_{\rm max})$ is derived from the assumption that the size distributions of both the dust populations can be described by a power law with a fixed exponent of $-3.5$  \citep[see \S 2.2 of][]{Vorobyov+2020}.
The term $\vec{f}_{\rm p}$ is the drag force per unit mass due to the back-reaction on the gas due to dust as mentioned earlier in Appendix \ref{sec:gas}. 
The drag force is calculated using the Henderson drag force coefficient \citep{henderson1976}, which is valid both in the Epstein and Stokes regime of dust dynamics \citep[see][for details]{Stoyanovskaya+2020,Vorobyov2023}.
We note that, following \citet{Zhu2012}, we have added a small pressure to the dust fluid, which is equal to 0.001\% that of the local gas pressure. This is done to smooth delta shocks that may appear in a zero-pressure dust fluid when dust-to-gas friction becomes small for large $\mathrm{St}$, but its effect on dust drift is negligible.

The evolution of the maximum radius that the grown dust can achieve obeys a continuity equation
\begin{equation}
{\partial a_{\rm max} \over \partial t} + ({\vec u}_{\rm p} \cdot \vec{\nabla}_p ) a_{\rm max} = \cal{D} \ ,
\label{eq:dustA}
\end{equation}
where the growth rate $\cal{D}$ represents the coagulation and can be written in the monodisperse dust growth approximation as 
\begin{equation}
\cal{D}=\frac{\rho_{\rm d} {\it v}_{\rm rel}}{\rho_{\rm s}} \ 
\label{eq:GrowthRateD}
\end{equation}
\citep{BirnstielEtal2012}.
Here, $\rho_{\rm d}$ is the total dust volume density and $v_{\rm rel}$ is the relative dust collision velocity derived following closed form expressions for turbulent eddies introduced by \citet{Ormel2007}
\begin{equation}
    v_{\rm{rel}} = \sqrt{{3 \alpha_{\rm visc} \over \mathrm{St}+\mathrm{St}^{-1}}} c_{\rm s}\, ,
    \label{turb_vel}
\end{equation}
where $\mathrm{St}$ is the Stokes number. This expression accurately captures the behavior of the dust collision velocity at small and large $\mathrm{St}$. 
When calculating the volume density of grown dust, we take into account dust settling by calculating the effective scale height of grown dust $H_{\rm d}$ via the corresponding gas scale height $H_{\rm g}$, $\alpha_{\rm visc}$ parameter, and the Stokes number ${\rm St}$ as 
\begin{equation}
H_{\rm d} = H_{\rm g}   \sqrt{\frac{\alpha_{\rm visc}}{\alpha_{\rm visc} + {\rm St}}}\, ,
\end{equation}
where the Stokes number is defined as
\begin{equation}
    {\rm St} = \frac{{\rm \Omega_{k} \rho_{\rm s} {a_{\rm max}}}}{\rho_{\rm g} {\rm c_s}} =\Omega_{k} t_s \, ,
    \label{eq:stokes}
\end{equation}
where $\Omega_{\rm K}$ is the Keplerian orbital frequency and $t_{\rm s}$ is the dust stopping time. 
Note that, dust particles with Stokes number ${\rm St} \ll 1$ adapt to
the gas velocity on time scales much shorter than the orbital time scale, while
the dust particles with Stokes number ${\rm St} \gg 1$ perform several
orbits before the drag forces significantly alter their velocity.
The scale height of small dust is assumed to be equal to that of the gas.

Furthermore, the maximum size of $a_{\rm max}$ is limited by the fragmentation barrier 
\begin{equation}
 a_{\rm frag}=\frac{2\Sigma_{\rm g}v_{\rm frag}^2}{3\pi\rho_{\rm s}\alpha_{\rm visc} c_s^2}
 \label{afrag}
\end{equation}
\citep{Birnstiel+2016},
where  $v_{\rm frag}$ is the fragmentation velocity, namely, a threshold
value of the relative velocity of dust particles at which collisions
result in fragmentation rather than coagulation.
In our model, $v_{\rm frag}$ is set to $5 \,{\rm m} \, {\rm s}^{-1}$ and $c_s$ is the local sound speed. 
The growth rate $\cal{D}$ is set to zero when $a_{\rm max}$ exceeds $a_{\rm frag}$. 
We note that if the fragmentation barrier is reached and dust growth halts $a_{\rm max} = a_{\rm frag}$, the local conditions in the disk can
change such that the value of fragmentation barrier decreases (for instance, if the gas density decreases or temperature rises).
If this occurs, $a_{\rm max}$ is adjusted accordingly to the new value of $a_{\rm frag}$. We note that the so-called drift barrier is accounted for self-consistently via the computation of the grown dust dynamics \citep[see further details in][]{Vorobyov+2024}.


\subsection{Magnetorotational instability, adaptive viscosity, and ionization fraction} \label{sec:viscosity_ion}
The MRI acts in the presence of the magnetic field and causes turbulence in the weakly ionized gas in the shearing Keplerian disk \citep{Balbus1998}. 
The effects of the turbulence generated by the MRI can be described by a viscosity \citep{ShakuraSunyaev1973}, and  
turbulent viscosity generated by the MRI is traditionally thought to play a major role for the mass and angular momentum transport in protostellar/protoplanetary disks.

The development of the MRI requires a certain ionization level and the Galactic cosmic rays are the primary source of ionization, which are external to the disk and penetrate the disk from both above and below. The disk in this picture is thought to be accreted through a magnetically layered structure \citep[][]{Gammie1996, Armitage+2001} as shown in Figure 5.2 of \cite{IDthesis2022}, wherein most accretion occurs via the MRI-active surface layers and a magnetically dead zone may be formed at the midplane if the total disk column density exceeds $\sim 200$~g~cm$^{-2}$. The effective $\alpha_{\rm visc}$ averaged over the disk vertical column may reach $10^{-3}-10^{-2}$ in disks with the sustained MRI to match the observed mass accretion rates \citep{1995ApJ...446..741B,1996ApJ...463..656S,Armitage+2001}.

This layered disk model has recently been brought into question, following observations of  efficient dust settling in protoplanetary disks \citep{Rosotti2023} and numerical magnetohydrodynamics simulations with nonideal MHD effects included \citep{BaiStone2013,Gressel2015}. According to these findings, the MRI may be effectively suppressed across most of the disk extent and we set $\alpha_{\rm visc}$ to a low value of $10^{-4}$, meaning that turbulent viscosity provides only residual accretion via a thin surface layer of the disk.

There are, however, situations when the MRI may still play an important role.   If the temperature of the inner disk rises above 1000~K, the alkali metals (e.g., potassium) become thermally ionized, providing sufficient ionization for the MRI activation across the entire vertical column of the disk, which in turn leads to an accretion burst.  
In the numerical setup of FEOSAD, we account for this possibility by considering the adaptive $\alpha$-parameter. The kinematic viscosity $\nu = \alpha_{\rm visc} c_s H_{\rm g}$ is parameterized using the usual \cite{ShakuraSunyaev1973}  prescription. 
To simulate accretion bursts, we redefine $\alpha_{\rm visc}$ as a density weighted average which follows:
\begin{equation}
\label{eq:alphaeff}
    \alpha_{\rm visc} = \frac{\Sigma_{{\rm MRI}} \, \alpha_{{\rm MRI}} + \Sigma_{{\rm dz}} \, \alpha_{{\rm dz}}}{\Sigma_{{\rm MRI}} + \Sigma_{{\rm dz}}}\, ,
\end{equation}
where $\Sigma_{{\rm MRI}}$ is the gas column density of the MRI-active layer and $\Sigma_{{\rm dz}}$ is that of the magnetically dead layer at a given radial distance, so that $\Sigma_{\rm g} = \Sigma_{{\rm MRI}} +\Sigma_{{\rm dz}}$. 
The values of $\alpha_{\rm MRI}$ and $\alpha_{{\rm dz}}$ in Equation~(\ref{eq:alphaeff}) correspond to the strength of the turbulence in the MRI-active layer and the MRI-dead  layer, respectively. 
Across the disk, the constant background turbulent viscosity $\alpha_{\rm MRI}$ is set equal to $10^{-4}$. However, if the thermal ionization exceeds $x_{\rm th}=10^{-10}$, which occurs during the MRI burst, $\alpha_{\rm MRI}$ is then set to 0.1. 
The latter value is motivated by three-dimensional numerical hydrodynamics simulations of the MRI triggering \citep{Zhu2020}. 
Because of the nonzero residual viscosity arising from hydrodynamic turbulence driven by the Maxwell stress in the active layer, a very small value of $10^{-5}$ is considered for $\alpha_{{\rm dz}}$ \citep[][]{OkuzumiHirose2011}. 
The depth of the dead zone in terms of the $\alpha_{\rm visc}$-parameter can be now
determined by the balance between $\Sigma_{\rm MRI}$ and $\Sigma_{\rm dz}$ using Eq. (\ref{eq:alphaeff}).
The value of $\Sigma_{{\rm MRI}}$ is not set to be constant as in many studies of the MRI bursts \citep{Bae2014,Kadam+2022}, but is defined as
\begin{eqnarray}
\label{eq:zeroDZ}
          \Sigma_{\rm MRI} &=& \Sigma_{\rm g}, \, \, {\rm if} \ \Sigma_{\rm g} < \Sigma_{\rm crit} \, , \\ 
     \Sigma_{\rm MRI} &=& \Sigma_{\rm crit},  \, \,    
    {\rm if} \ \Sigma_{\rm g} \geq \Sigma_{\rm crit} \, , 
\end{eqnarray}
where the critical gas surface density $\Sigma_{{\rm crit}}$  for the MRI development is obtained by equating the timescale of MRI growth to the timescale of MRI damping due to Ohmic dissipation \citep[see details in][]{Vorobyov+2020}, and is expressed as 
\begin{equation}
    \Sigma_{{\rm crit}} = \left[\left(\frac{\pi}{2}\right)^{1/4}\frac{c^2m_{\rm e}\langle\sigma v\rangle_{\rm en}}{e^2}\right]^{-2}B_z^2 H_{\rm g}^3 x^2 \ ,
    \label{eq:DZ}
\end{equation}
where $e$ is the charge of the electron, $m_{\rm e}$ is mass of the electron, $\langle\sigma v\rangle_{\rm{en}}=2\times 10^{-9}\,\mbox{cm}^3\,\mbox{s}^{-1}$ is the slowing-down coefficient \citep{spitzer1978} for the electron-neutral collisions, and $c$ is the speed of light. 

The ionization fraction $x=n_e/n_n$ is determined from the balance of collisional, radiative recombination, and recombination on dust grains. This is expressed as
\begin{equation}
(1-x)\xi = \alpha_{\rm{r}} x^2n_{\rm{n}} + \alpha_{\rm{d}} xn_{\rm{n}} \, ,
\label{eq:ion}
\end{equation}
where $\xi$ is the ionization rate that is composed of a cosmic-ray ionization rate and the ionization rate by radionuclides \citep{UmebayashiNakano2009}, $n_{\rm n}$ is the number density of neutrals, $\alpha_{\rm{d}}$ is the total rate of recombination onto the dust grains, and $\alpha_{\rm{r}}$ is the radiative recombination rate having a form $\alpha_r = 2.07 \times 10^{-11} \, T^{-1/2} \, {\rm cm}^3 {\rm s} ^{-1}$ \citep{spitzer1978}. 
In the regions, where the gas temperature exceeds several hundred Kelvin, an additional term is added to the ionization fraction $x$ from Eq. \ref{eq:ion}, which is the thermal ionization calculated by considering the ionization of potassium, the metal with the lowest ionization potential. 
The cosmic abundance of potassium set to $10^{-7}$ for these calculations. For further details, including the calculation of the total recombination rate on dust grains $a_{\rm d}$, we refer to  \citet{Vorobyov+2020}. 
Equation~(\ref{eq:zeroDZ}) demonstrates that a sharp increase in $\Sigma_{\rm crit}$ triggers the burst if the ionization
fraction $x$ experiences a sharp rise as well.

\subsection{Initial and Boundary Conditions}\label{sec:ICs}
The initial profile of the gas surface density $\Sigma_g$ and angular
velocity $\Omega$ of the prestellar core has the following form: 
\begin{equation}
\Sigma_{\rm g} = \frac{r_0 \Sigma_0}{\sqrt{r^2+r_0^2}} \ ,
\end{equation}
\begin{equation}
    \Omega = 2\Omega_0 \left(\frac{r_0}{r}\right)^2 \left[\sqrt{1+\left(\frac{r}{r_0}\right)^2} -1 \right] \ ,
\end{equation} 
consistent with an axially symmetric core collapse  \citep{Basu1997}.
Here, $\Sigma_0$ and $\Omega_0$
are the gas surface density and angular speed at the center of the core, $r_0 = \sqrt{A} c_s^2/ (\pi G \Sigma_0)$ is the radius of the central plateau, where $c_s$ is the local sound speed in the core, $r$ is the radial distance from the center. 
The dimensionless parameter $A$ corresponds to the density perturbation and it is set to $2$ that makes the core unstable to collapse \citep[see][for more details]{Vorobyov+2020}. 
The initial dust-to-gas ratio is set to $1:100$, where all the dust is in the form of small sub-micron dust grains that are fully coupled with the gas component. 
The gas temperature of the initial prestellar core is set to $T_{\rm}= 20$ K and a uniform background vertical magnetic field of strength is taken as $B_0 = 10^{-5} \, {\rm G}$. 
The spatially uniform mass-to-magnetic flux ratio $\lambda= (2 \pi \sqrt{G})\Sigma_{\rm g}/B_z$ is set equal to 10.0, and it stays constant during the entire disk evolution in the adopted ideal MHD limit. 
The initial prestellar cores with a supercritical mass-to-flux ratio, i.e., $\lambda >1$, are understood to form through ambipolar diffusion (neutral-ion drift) driven gravitational collapse.


The inner boundary in FEOSAD represents a circular sink cell with a radius of 0.52~au.  Placing the inner boundary much farther out (at several au) could eliminate the part of inner disk that may be dynamically important since it is where the GI-induced MRI outbursts take place.  The mass exiting the inner disk is divided between the sink cell, the star, and the jet, with the ratio set to $5\%:85.5\%:9.5\%$. It means that most of the matter quickly lands on the star after crossing the sink-disk interface, but smaller amounts are ejected with the jets and kept in the sink cell. In this realization, the sink cell is characterized by its own surface density of gas and dust, which allows setting an inflow-outflow boundary condition, wherein the matter is allowed to flow freely from the disk to the sink cell and vice versa.  This type of boundary condition allows reducing an artificial drop in the gas density at the inner disk edge, which often forms in outflow-only boundary conditions because of the lack of compensating back flows during wave-like motions triggered, for example, by the spiral density waves or Coriolis force. The angular velocity at the inner boundary is extrapolated according to the Keplerian pattern of rotation and the radial velocity is set equal to the corresponding value of the innermost active disk cell. The flow of matter to and from the sink cell also carries magnetic flux, hence, the inner boundary condition also modifies the vertical component of magnetic field $B_z$ based on the amount of magnetic flux transported. The outer boundary of the computational domain is taken as standard free outflow, where the material is only allowed to leave (no inflow) the computational domain. More details can be found \citet{Vorobyov+2018} and \citet{Vorobyov+2020b}.

\section{Propagation of an outside-in outburst}

\begin{figure*}[htb!]
\centering
\begin{minipage}{\linewidth}
\centering
\includegraphics[width=\linewidth]{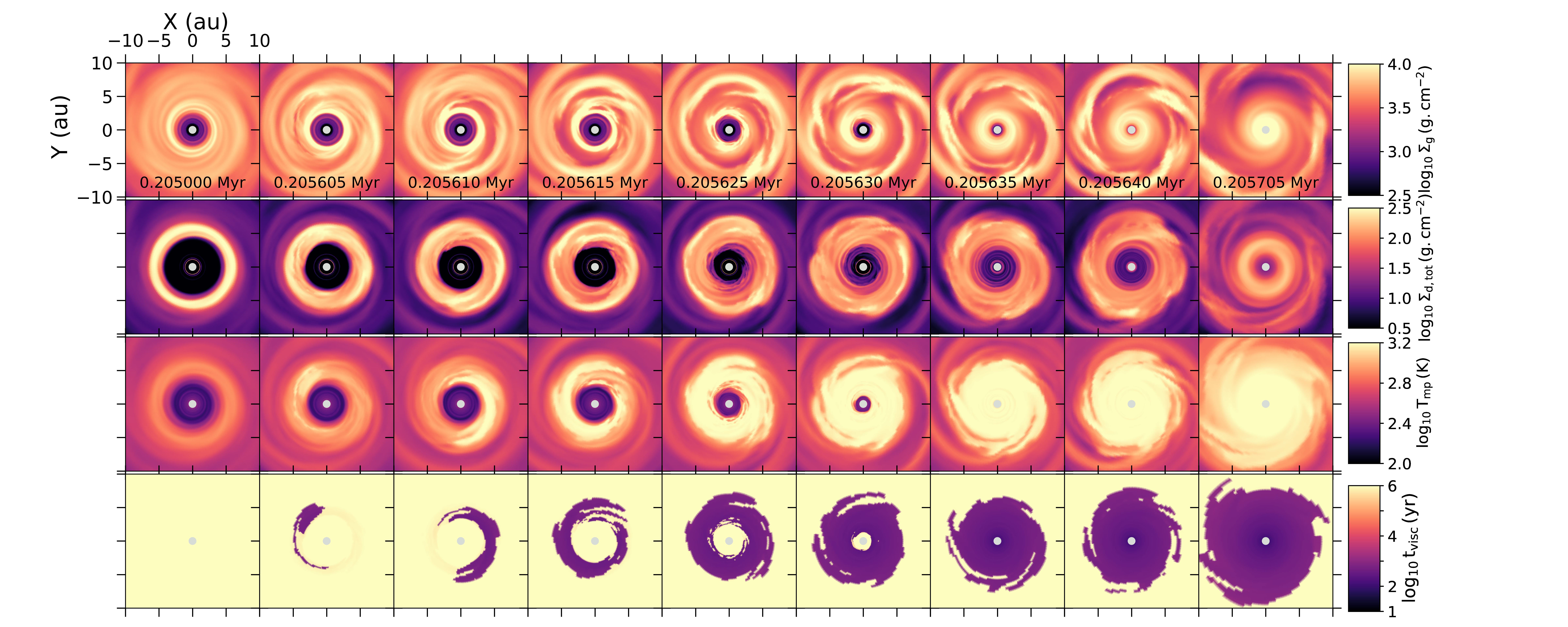}
\end{minipage}
\hfill
\begin{minipage}{\linewidth}
\hspace{-1cm}
\centering        \includegraphics[width=0.8\linewidth,height=4cm]{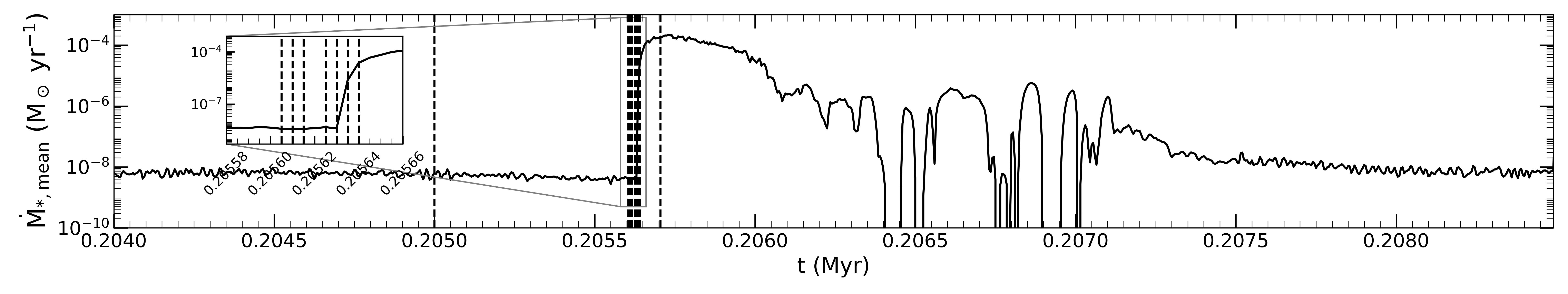}
\end{minipage}
\caption{The progression of an outside-in MRI outburst as obtained from \simname{model-YIMS}. From top to bottom: Evolution of the gas surface density ($\Sigma_{\rm g}$) and total dust surface density ($\Sigma_{\rm tot}$) in units of $\log_{10} \, {\rm g \, {cm}^{-2}}$, midplane temperature ($\Temp$) in units of $\log_{10} \, {\rm (K)}$, viscous timescale ($t_{\rm visc}$) in units of  $\log_{10} \, {\rm yr}$ over a region of $20 \times 20 \, {\rm au}^2$ in the midplane of the disk. The bottom panel represents the temporal evolution of mass accretion onto the star at the sink-disk interface as a single burst develops and decays, with a focus that zooms in on the rise of the outburst. The vertical dashed lines mark the time instances corresponding to the above 2D intensity maps.}
\label{fig:outsideinburst}
\end{figure*}

\begin{figure*}[htb!]
   \centering
\includegraphics[width=\linewidth]{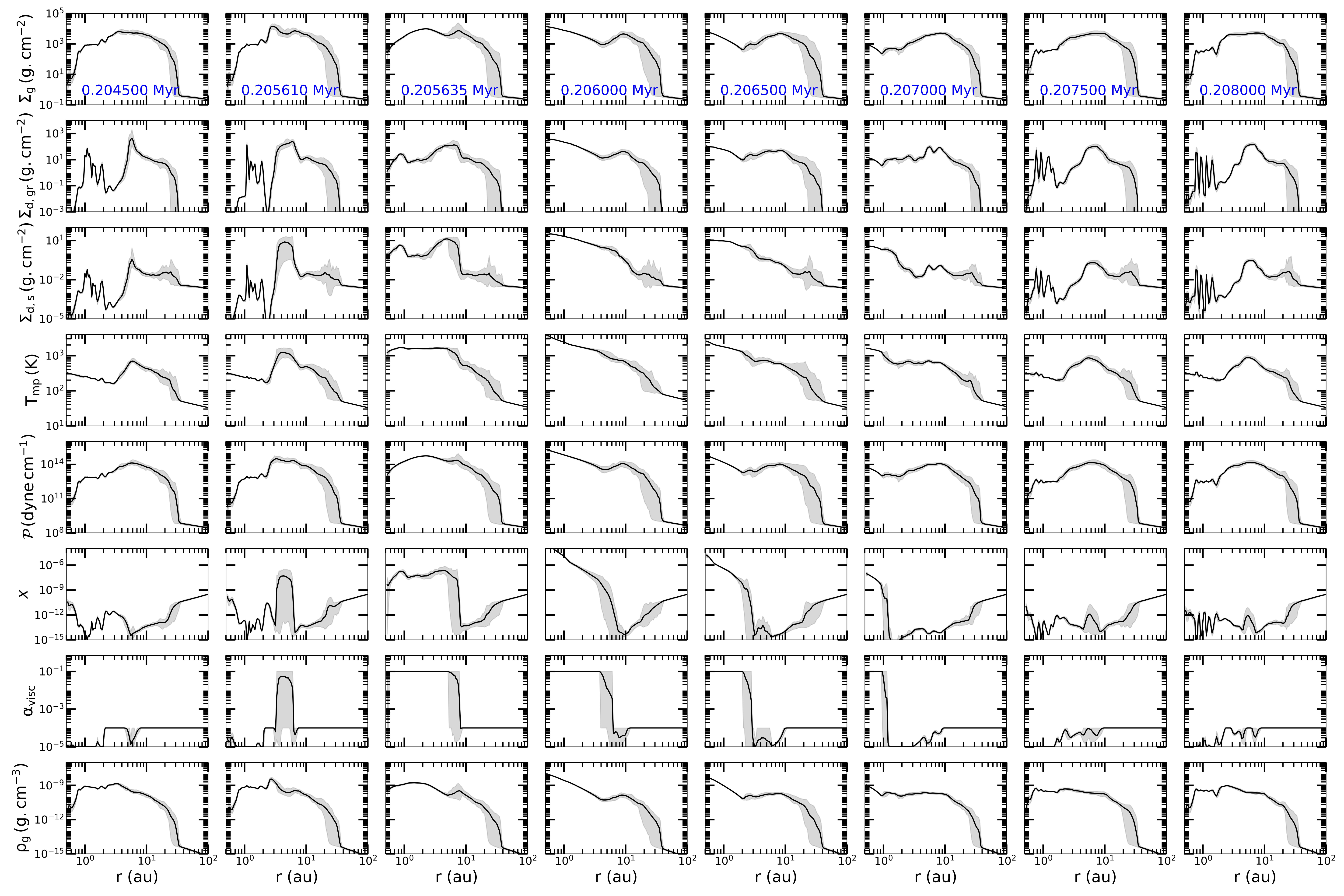}
\caption{The azimuthally averaged radial profiles of the disk characteristics as at the respective time instances 
as shown in the intensity map of Figure \ref{fig:gasdust2DEvol} for \simname{model-YIMS}. From top to bottom: evolution of the gas surface density $\Sigma_{\rm g}$, grown dust ($\Sigma_{\rm d,gr}$) and small dust surface density ($\Sigma_{\rm d,s}$), midplane temperature $\Temp$ (in units of ${\rm K}$), vertically integrated pressure ($\mathcal{P}$), ionization-fraction  ($x$), and turbulent viscosity ($\alpha_{\rm visc}$), gas volume density ($\rho_{\rm g}$) within the disk over a region of 100~au. 
The gray shaded area around the respective thick black curves show the azimuthal scatter at a given radial distance.}
\label{fig:gasdust1DEvol}
\end{figure*}

Figure \ref{fig:outsideinburst} showcases the propagation of an outburst in an ``outside-in" fashion. 
It illustrates a 2D spatial representation of the disk characteristics within the inner region of PPD to understand how a single outburst develops with a focus that zooms in on the rise of the outburst. 
As the burst progresses, the structure in the inner region of PPD is characterized by a spiral pattern in gas ($\Sigma_{\rm g}$) and in the total dust surface density ($\Sigma_{\rm d,tot}$). 
The midplane temperature, $T_{\rm mp}$ rises to over a few
thousand Kelvin during the outburst.
The viscous timescale, which measures the time it takes for matter to be transported from the disk to the star, drops by roughly 5 orders of magnitude within the innermost region of the PPD, implying the rapid transport of the inner disk gas onto the star owing to sharply increased turbulent
viscosity.  
An episodic outburst can potentially crush the accretion funnels, leading to accretion occurring through a boundary layer. 
The mean and burst accretion rates for YIMSs are higher than those of low-mass stars by up to an order of magnitude \citep[for low-mass stars, see further in][etc.]{Vorobyov+2020,Kadam+2022}.

Figure \ref{fig:gasdust1DEvol} shows the azimuthally averaged radial profiles of disk variables for a gas-dust+MHD disk at specific time instances as a single outburst develops and decays, similar to those epochs shown in Figure \ref{fig:gasdust2DEvol}. 
In our GI-dominated model, a spatially varying rate of mass transport due to gravitational torques results in a bottleneck and dead zone in the inner disk. 
This leads to spatially varying density gradients, yielding local pressure maxima that cause the dust to concentrate locally into rings. 
The increased dust concentration within the massive dust rings makes the region optically thicker, gradually warming the nearby gas and aiding the initiation of MRI once the midplane temperature rises from a few hundred Kelvin in the pre-outburst stage to over a few thousand Kelvin during the outburst.

\section{Tracer of dead zone within PPD}
\begin{figure*}
\includegraphics[width=\linewidth]{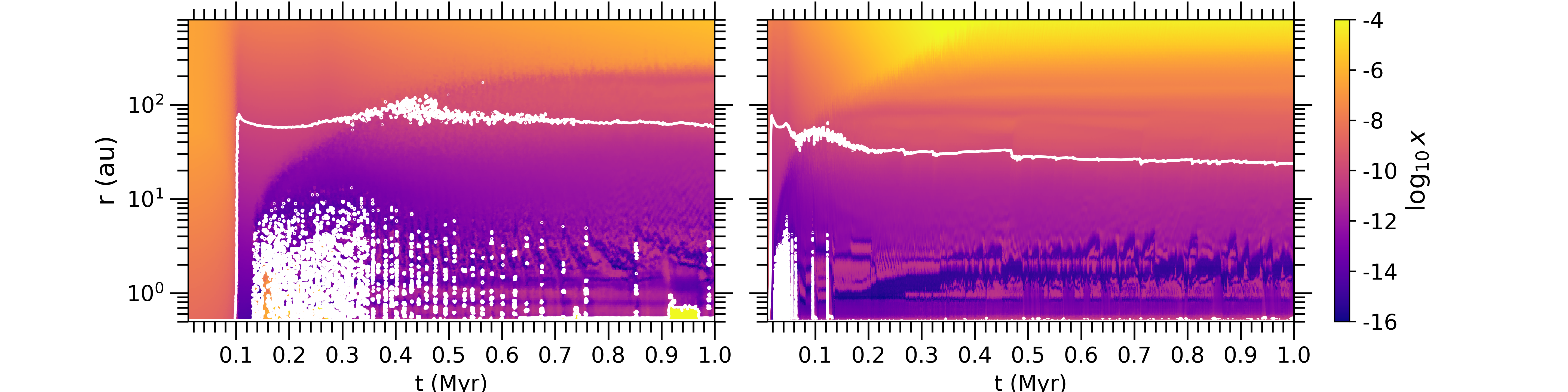}
\caption{Spacetime map of azimuthally averaged ionization fraction (in logarithmic units) overlaid with the contour line $x=10^{-10}$ representing the dead zone boundary. The left and right panel show the cases for \simname{model-YIMS} and \simname{model-LMS}, respectively.}
\label{fig:xIon}
\end{figure*}
Figure \ref{fig:xIon} shows the spacetime diagram of the azimuthally averaged ionization-fraction. For \simname{model-YIMS}, GI-sustained MRI-triggered bursts occur within 10-12~au, whereas, for \simname{model-LMS}, the bursts are confined within 5-6~au. 
In these considered
cases, the outer boundary of the dead zone represented by the contour line corresponding to $x=10^{-10}$ lies just outside the dust rings. 
The extent of the dead zone in \simname{model-YIMS} is greater
than in \simname{model-LMS}. 
By the timeline of 1~Myr, the dead zone boundary is located at $r\approx$~60~au for \simname{model-YIMS} and  $r\approx$~25~au for \simname{model-LMS}.


\section{Dust sublimation}
\label{sec:DustEvapApp}

\begin{figure*}[htb!]
    \centering
    \begin{minipage}{\linewidth}
       \centering
       \includegraphics[width=0.5\linewidth]{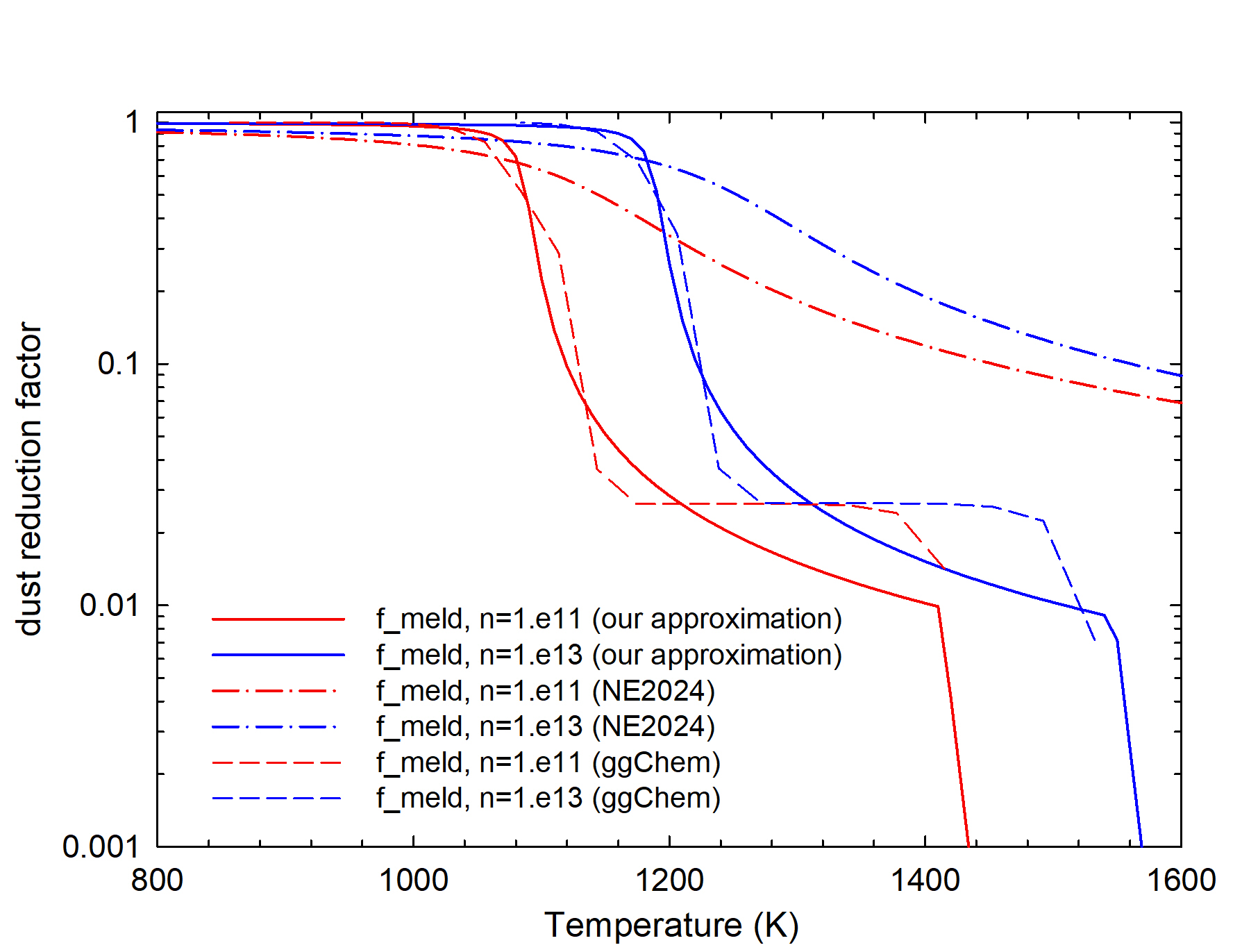} 
       \caption{Dust depletion factor as a function of local temperature and density. The dashed lines show the values obtained from the ggChem model \citep{2018A&A...614A...1W}, while the dash-dotted and solid lines provide the fits from \citet{NayakshinElbakyan2024} and present work.} 
\label{fig:DustRedFact}
    \end{minipage}
    \hfill
    \vspace*{0.5cm}
    \begin{minipage}{\linewidth}
        \centering
\includegraphics[width=\linewidth]{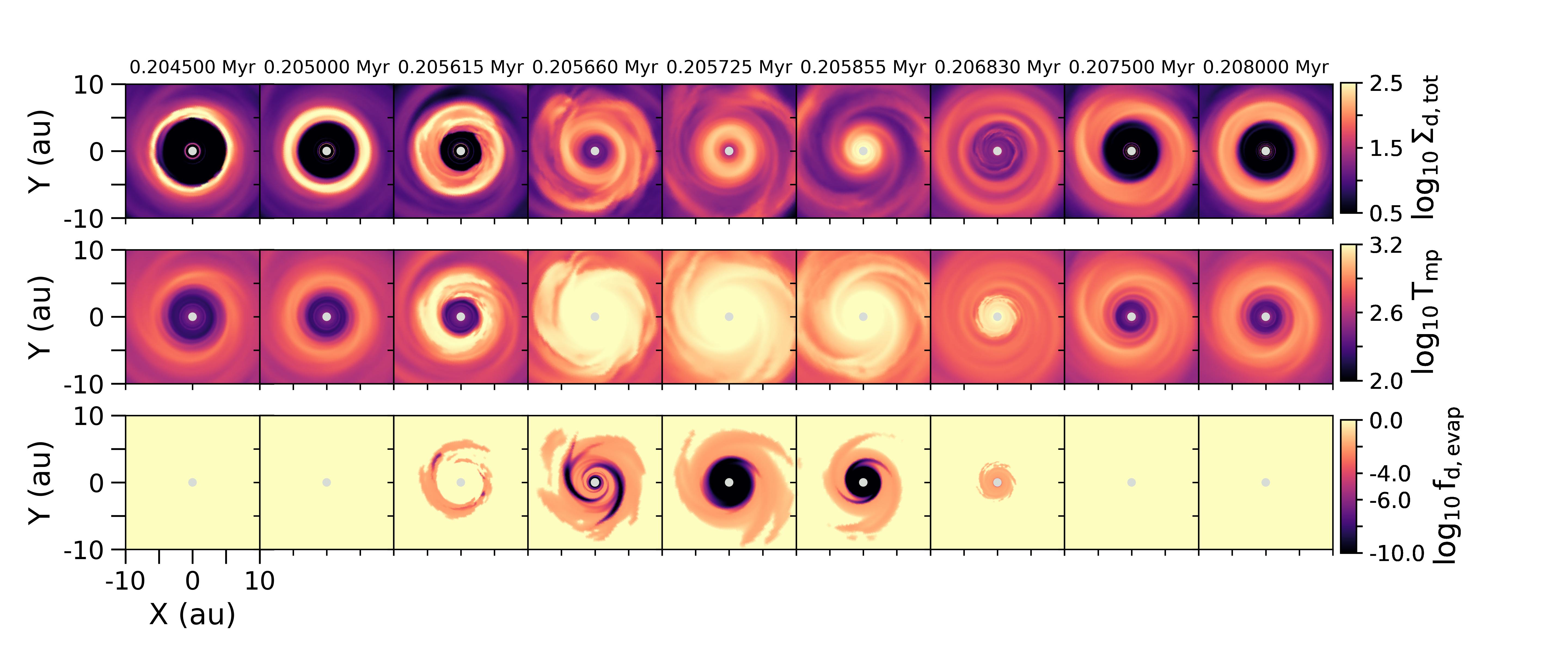}
    \end{minipage}
    \hfill
    \begin{minipage}{\linewidth}
         \hspace{-1cm}
        \centering        \includegraphics[width=0.8\linewidth,height=3cm]{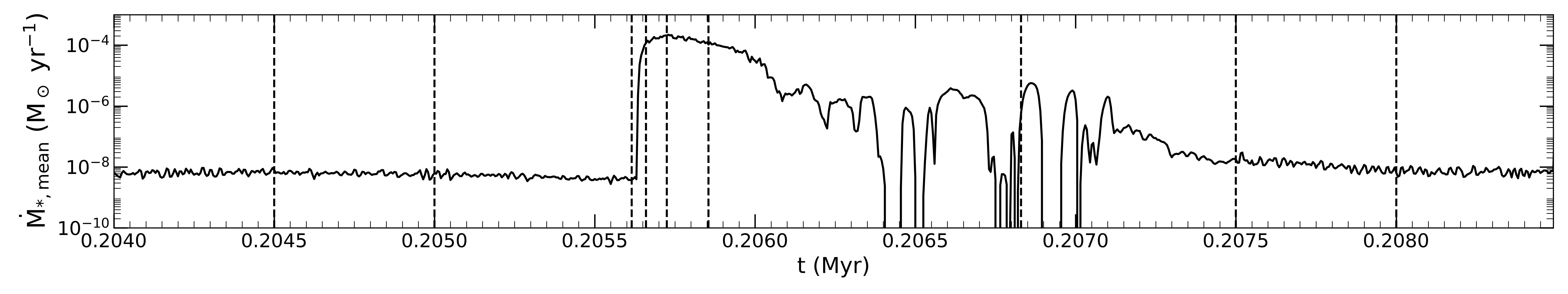}
        \caption{MRI outburst progression and decay as presented in 2D maps of total dust surface density, midplane temperature, and fraction of dust reduction due to dust evaporation as obtained from \simname{model-YIMS}. 
The bottom panel shows the corresponding temporal evolution of mass accretion rate onto the star at the sink-disk interface and the vertical dashed lines mark the time instances corresponding to the 2D intensity maps.} 
\label{fig:2DTmeltburstv2}
    \end{minipage}
\end{figure*}

Dust temperature in the innermost disk regions may rise above the threshold for dust sublimation. This effect is considered in FEOSAD via the dust opacity \citep{Semenov+2003}, which sharply decreases above this threshold. Dust, however, is not physically removed from the system. To evaluate the effect of dust sublimation, we post-processed our model searching for disk regions where dust can sublimate. To establish the dust evolution under extreme temperatures,
we used  the astrochemistry code \href{https://github.com/pw31/GGchem}{ggCHEM}, which simulates phase transitions based on equilibrium chemistry using detailed thermochemical data of heavy element condensation and sublimation within the innermost region of a PPD \citep{2018A&A...614A...1W}. The dashed lines in Figure~\ref{fig:DustRedFact} display the dust depletion factor $f_{\rm evap}$ as a function of temperature and density, namely, the factor showing the decrease in the local mass of dust as the temperature rises and dust sublimates. Dust experiences two major episodes of sublimation: around $1100-1200$~K when silicate and carbon refectory species sublimate and around $1400-1500$~K when most refractory species (for example, $\mathrm{Al_2O_3}$) sublimate. Above 1600~K, dust is basically nonexistent.

For comparison purposes, we also plot in Figure~\ref{fig:DustRedFact} the dust depletion factor from \citet{NayakshinElbakyan2024}, defined following \citep{Kuiper+2010} as
\begin{equation}
    f_{\rm evap} = \frac{1}{2}-\frac{1}{\pi} \, {\rm arctan} \left( \frac{{T-T_{\rm evap}}}{100 \, {\rm K}} \right),
    \label{eq:dust-evap-NE1}
\end{equation}
where 
\begin{equation} 
    T_{\rm evap} = 2000 \, {\rm K} \,\,\, \rho_{\rm g}^{0.0195},
    \label{eq:dust-evap-NE2}
\end{equation}
and the volume density of gas $\rho_{\rm g}$ (in g~cm$^{-3}$) is obtained from the corresponding gas surface density and the gas disk scale height with the assumption of the local hydrostatic equilibrium.
As can be seen, Equations~(\ref{eq:dust-evap-NE1}) and (\ref{eq:dust-evap-NE2}) reproduce only the basic behavior of the more accurate ggCHEM model. 

We update these equations to achieve a better agreement as follows
\begin{eqnarray}
    f_{\rm melt} &=& \frac{1}{2}-\frac{1}{\pi} \, {\rm arctan} \left( \frac{{T-T_{\rm melt}}}{10 \, {\rm K}} \right), \,\, \mathrm{if} \, T<T_{\rm evap} \label{eq:dustEvapOur1} \\
    f_{\rm evap} &=& f_{\rm melt} \times \exp \left(-\frac{{T - T_{\rm evap}}}{10 \, {\rm K}} \right), \,\, \mathrm{if} \, T>T_{\rm evap},
\end{eqnarray}
where 
\begin{equation} \label{eq:Tmelt}
    T_{\rm melt} = 1900 \, {\rm K} \, \rho_{\rm g}^{0.0195}  \ ,
\end{equation}
and
\begin{equation} 
    T_{\rm evap} = 2500 \, {\rm K} \, \rho_{\rm g}^{0.02} .
    \label{eq:dustEvapOur2}
\end{equation}
Equations~(\ref{eq:dustEvapOur1})-(\ref{eq:dustEvapOur2}) can reproduce the step-like decrease in $f_{\rm evap}$ with rising temperature to a better degree of accuracy.

We found that the dust-evaporation prone zones are sporadic and mostly coexist with the appearance of MRI-outbursts until the end of the simulation. 
To provide more details on dust evaporation during MRI bursts, we analyze it as an outburst emerges and decays, as shown in Figure \ref{fig:2DTmeltburstv2}. The dust depletion fraction is shown for the respective time instances along with the total dust surface density. 
Clearly, dust is supposed to sublimate during the bursts in an inner region up to 5-7 au. We note that some of this to-be-sublimated dust may end up being accreted during the burst. The effect of dust sublimation should be considered in future simulations of accretion bursts to assess the relative fractions of accreted and sublimated dust.

\bibliography{myref}{}
\bibliographystyle{aasjournal}

\end{document}